\documentclass[a4paper,fleqn]{cas-sc}
\usepackage[justification=centering]{caption}
\usepackage[numbers]{natbib}

\usepackage{graphicx}
\usepackage[section]{placeins}

\usepackage{color}
\usepackage{float}
\newtheorem{definition}{Definition}  

\usepackage{booktabs}

\usepackage{algorithm}

\begin{document}

\shorttitle{Contrast Pattern Mining: A Survey} 

\shortauthors{Yao Chen et~al.}

\title [mode = title]{Contrast Pattern Mining: A Survey}                      

\author[1]{Yao Chen}
\ead{csyaochen@gmail.com}
\address[1]{College of Cyber Security, Jinan University, Guangzhou 510632, China}

\author[1,2]{Wensheng Gan}
\cormark[1]
\ead{wsgan001@gmail.com}
\address[2]{Pazhou Lab, Guangzhou 510330, China}
\cortext[cor1]{Corresponding author}

\author[1]{Yongdong Wu}
\ead{wuyd175@gmail.com}

\author[3]{Philip S. Yu}
\ead{psyu@uic.edu}
\address[3]{Department of Computer Science, University of Illinois at Chicago, Chicago, USA}

\begin{abstract}	
  Contrast pattern mining (CPM) is an important and popular subfield of data mining. Traditional sequential patterns cannot describe the contrast information between different classes of data, while contrast patterns involving the concept of contrast can describe the significant differences between datasets under different contrast conditions. Based on the number of papers published in this field, we find that researchers' interest in CPM is still active. Since CPM has many research questions and research methods. It is difficult for new researchers in the field to understand the general situation of the field in a short period of time. Therefore, the purpose of this article is to provide an up-to-date comprehensive and structured overview of the research direction of contrast pattern mining. First, we present an in-depth understanding of CPM, including basic concepts, types, mining strategies, and metrics for assessing discriminative ability. Then we classify CPM methods according to their characteristics into boundary-based algorithms, tree-based algorithms, evolutionary fuzzy system-based algorithms, decision tree-based algorithms, and other algorithms. In addition, we list the classical algorithms of these methods and discuss their advantages and disadvantages. Advanced topics in CPM are presented. Finally, we conclude our survey with a discussion of the challenges and opportunities in this field.
\end{abstract}


\maketitle

\section{Introduction}   \label{sec:introduction}

With the rapid development of technologies such as the Internet of Things and artificial intelligence, the volume of data has explosive growth, with data types ranging from single structured data to diverse unstructured data and semi-structured data. According to a Cisco report, people, machines, and things will generate about 850 zettabytes (ZB) of data by 2021 \cite{cisco2021cisco}. The amount of information generated but not stored is two orders of magnitude higher than the amount of information that ends up being stored, and there is also the possibility of data graves (data being stored but never analyzed). This means that while the raw data is rich, valuable knowledge or information is scarce. In order to discover the valuable knowledge and information behind the data, data mining and data analysis technology have rapidly developed and a mature theoretical knowledge and technical system has been formed \cite{chen1996data,fournier2022pattern}. Data mining, also known as knowledge discovery, combines concepts and techniques from different fields and focuses on extracting knowledge or information from large amounts of application data \cite{gan2017data,gan2019survey}. Data analysis can be divided into two categories: prediction task and description task \cite{ventura2018supervised}. Predictive tasks, which aim to predict interesting information and trends, are studied by the machine learning community. While descriptive tasks, which aim at finding understandable patterns, are the research tasks of the data mining community  \cite{luna2019frequent,gan2021survey}. Contrast pattern mining (CPM) \cite{dong2012contrast} discussed in this article is a typical descriptive task. Contrast patterns describe significant differences between datasets with different contrast conditions.

CPM aims to find interesting contrast patterns in the data. This data can be classified into three categories: structured data, unstructured data, and semi-structured data. Structured data includes sequence data, graph data, web data, etc. Semi-structured data includes XML, JSON, etc. Unstructured data includes text data, music data, image data, etc. Since structured data management is basically in the form of a database and structured data is easy to process and statistically analyze. The type of data analyzed in this article is structured data. The core idea of CPM analysis data is contrast. Contrast is one of the most important means of analysis. The contrast analysis approach has the advantage of evaluating different options, helping to understand the problem, avoid potential risks, and choose the best solution. The advantage of using contrast to analyze problems is that the best solution can be found. In addition, contrast can be used to find connections between datasets and to help understand and solve problems. Comparisons involving two or more datasets are referred to as comparisons under statically defined conditions. Such datasets can be different subsets of a particular dataset or from different categories or time periods. On the contrary, comparisons where only one dataset is required for the comparison are referred to as comparisons under dynamically defined conditions. Contrast patterns describe the similarities and differences between datasets. CPM allows users to obtain interesting trends over different time, space, and classification or other contrast conditions. Building interpretable and accurate classifiers is also a classical application of CPM. CPM has become an important subfield of data mining that is effective in solving real-world problems, and researchers have used CPM to solve many practical problems with a wide range of CPM applications \cite{dong2012contrast}. For example, in the field of network traffic analysis, CPM can detect anomalous activities and identify changes in the system \cite{alipourchavary2020improving}.

\begin{figure}[h]
	\centering
	\includegraphics[height=5cm]{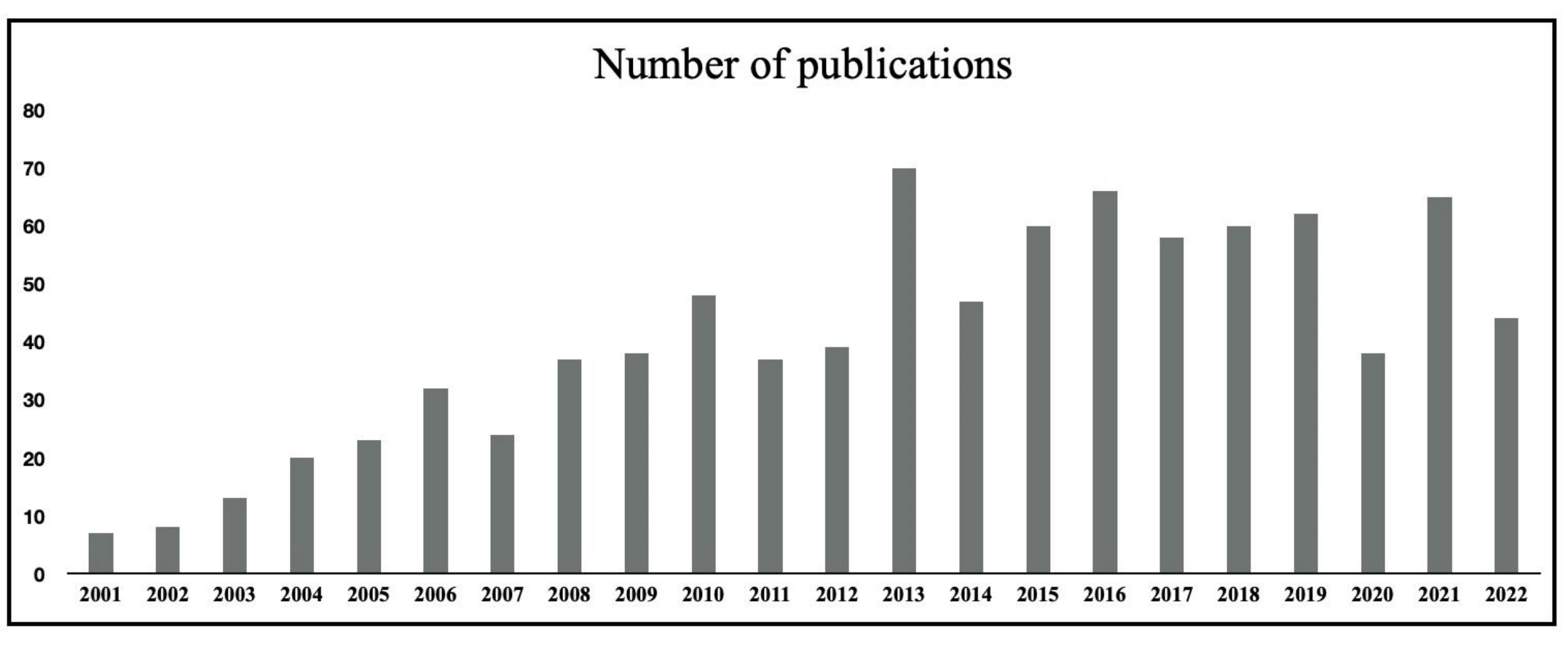}
	\caption{Statistics of papers retrieved on DBLP for contrast pattern mining.} \label{Fig 1}
\end{figure}

We used the DBLP website to count the number of papers published in the field of CPM for each year. The starting year of the search is 2001. The search phrase is "contrast|emerging|discriminative pattern". Figure 1 shows the results of the statistics. According to Figure \ref{Fig 1}, it is clear that the research interest in CPM remains active. Since CPM has many research questions and research methods. It is difficult for new researchers in the field to understand the general situation of the field in a short period of time. This article aims to help new researchers to have a general understanding of the basics of this research direction, current methods, advanced topics, and research opportunities. Therefore, the purpose of this article is to provide a detailed review and summary of the current state of CPM, attempting to provide an in-depth overview of the latest technologies, advanced topics, and challenges in this field.

\begin{itemize}
	\item  We provide an up-to-date comprehensive description of CPM in detail, including types of contrast pattern, mining strategies, methods of assessment of discriminative power, and applications.

	\item  We summarize the methods of CPM, including border-based, tree-based, evolutionary fuzzy systems-based, decision tree-based, and other algorithms. Use tables to help readers understand its development.

	\item  We review advanced topics in CPM in recent years, including class imbalance problems, fuzzy comparison models, CPM in big data environments, and data visualization.

	\item  Finally, there are still many interesting research questions in this research area, so we discuss several important issues and research opportunities in CPM.
\end{itemize}

The rest of this article is organized as follows. The following section introduces some basic concepts. In Section 3, the classification of CPM algorithms is introduced. In Section 4, we discuss the advanced topics of CPM. In Section 5, we describe the future directions and challenges of CPM. Finally, our conclusions are given in Section 6.

\section{Basic Concept: Contrast Pattern Mining}

\subsection{Preliminary and Types of Contrast Patterns}

\begin{figure}[h]
	\centering
	\includegraphics[width=0.6\columnwidth]{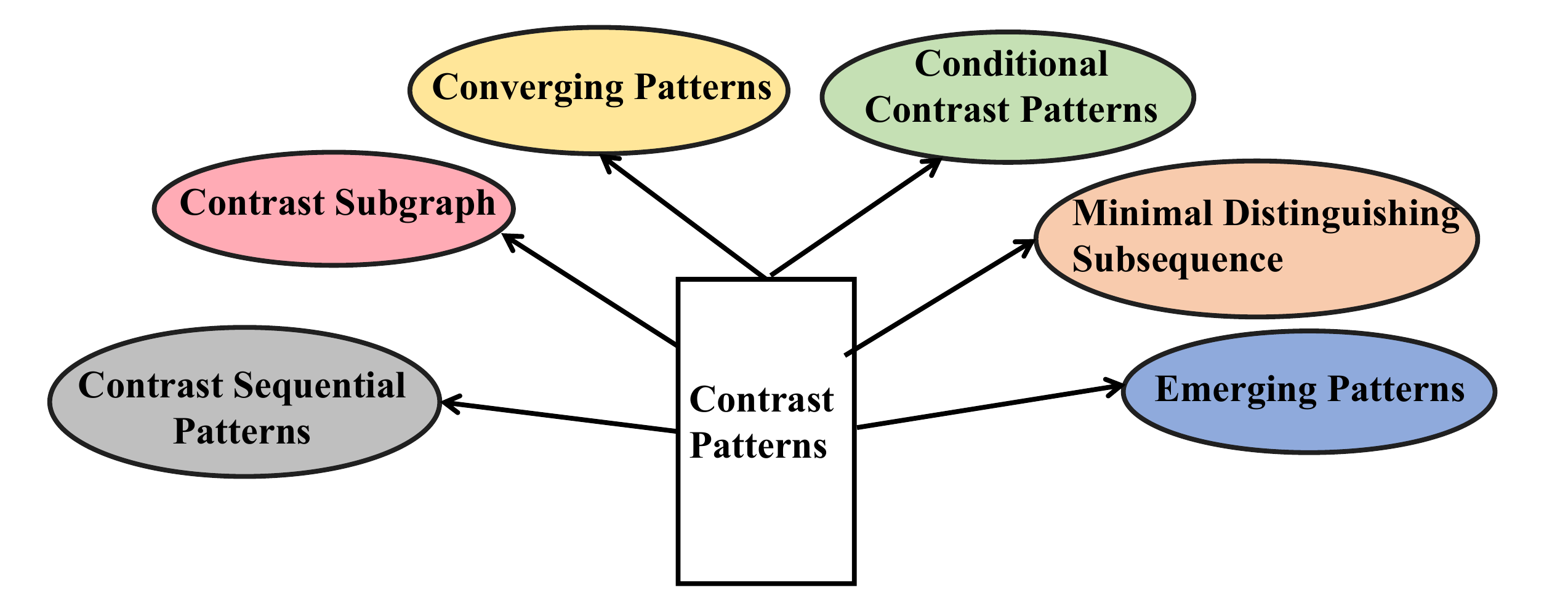}
	\caption{Contrast pattern family tree}
	\label{fig:CPFT}
\end{figure}

Table  \ref{Symbols} describes some of the basic concepts. Because of the diversity of contrast patterns. We made  a family tree of contrast patterns to introduce the common types of contrast patterns, as shown in Figure \ref{fig:CPFT}. The purpose of this section is to introduce the patterns mentioned in Figure \ref{fig:CPFT}.

\begin{definition}[Contrast Sequential Patterns, CSP \cite{wu2019mining,wu2021top,zheng2016effective}]
	\rm CSP is defined as a pattern that occurs frequently in one sequence dataset but not in the others. Before judging the pattern as CSP, we need to understand the concept of growth rate in advance. Given two sequence datasets: $D_1$ belongs to class $c_1$ and $D_2$ belongs to class $c_2$. The growth rate from $D_2$ to $D_1$ of a sequential pattern $X$ is denoted as:
\begin{equation}
	\label{GR}
$$
GR_{c_1}(X)= \left\{ \begin{array}{ll}
	\infty ,& \textrm{if sup($X$, $c_2$) = 0 \& sup($X$, $c_1$) $\neq$ 0}\\
	\frac{sup(X, c_1)/|D_1|}{sup(X, c_2)/|D_2|} ,& \textrm{otherwise}
\end{array} \right.
$$
\end{equation}

Similarly, the growth rate from $D_1$ to $D_2$ of a sequential pattern $X$ is denoted as: 
\begin{equation}
$$
GR_{c_2}(X)= \left\{ \begin{array}{ll}
	\infty ,& \textrm{if sup($X$, $c_1$) = 0 \& sup($X$, $c_2$) $\neq$ 0}\\
	\frac{sup(X, c_2)/|D_2|}{sup(X, c_1)/|D_1|} ,& \textrm{otherwise}
\end{array} \right.
$$
\end{equation}

The contrast rate of $X$ is denoted as:
\begin{equation}
$$CR(X) = \begin{cases}

\infty ,& \textrm{if GR$_{c_1}(X)$ = 0 or GR$_{c_2}(X)$ = 0}\\
	\max\{GR_{c_1}(X),GR_{c_2}(X)\}, &  \textrm{otherwise} \\
\end{cases}$$
\end{equation}

A sequence $s$ in a sequential database is said to be a CSP \cite{zheng2016effective} if its contrast rate is no less than the given CR threshold \textit{mincr}: CSP $\gets$ \{s | cr(s) $\ge$ \textit{mincr}\}. Unlike frequent sequential pattern mining \cite{wu2021top}, contrast sequential pattern mining (CSPM) \cite{wu2019mining} can discover the characteristics of different classes in sequence datasets, which has been widely used in sequential data analysis, such as protein/DNA dataset analysis, anomaly detection, and customer behavior analysis.
\end{definition}

\begin{table}[H]
	\caption{Summary of symbols and their explanations}
	\label{Symbols}
	\begin{tabular}{llll}  
		\toprule   
		\textbf{Symbol} & \textbf{Definition}  \\  
		\midrule  
		$I$ & A set of $m$ items, $I$ = \{ $i_1$, $i_2$, ..., $ i_m$\}.   \\ 
		itemset & An itemset is a subset of $I$. \\
		$s$ & A sequence  $s$ is an ordered list of itemsets. \\
		$D$ & A database $D$ with a set of items. \\  
		$sup(p, D)$ & The support of a pattern $p$ in a database $D$.\\
		CP & Contrast pattern.  \\  
		FP & Frequent pattern. \\
		CCP & Conditional contrast pattern.\\
		CSP & Contrast sequential pattern.\\
		EP & Emerging pattern.\\
		CPM & Contrast pattern mining.\\  
		FPM & Frequent pattern mining.\\
		CCPM & Conditional contrast pattern mining.\\
		CSPM & Contrast sequential pattern mining.\\
		EPM & Emerging pattern mining.\\
		MDS & Minimal distinguishing subsequence.\\
		MCP & multivariate contrast patterns.\\
		ZBDD & Zero-suppressed binary decision diagram.\\
		GEP & Gene expression programming.\\
			
		\bottomrule  
	\end{tabular}
\end{table}

\begin{definition}[Contrast Subgraph \cite{gionis2016bump,jiao2020sub,ting2006mining,wang2008spatial,yang2018mining}] 
	\rm Suppose there are two graph datasets $G_a$ = \{$G_{a1}$, $G_{a2}$, . . ., $G_{ax}$\} and $G_b$ = \{$G_{b1}$, $G_{b2}$, . . ., $G_{by}$\}. Subgraph $C$ is considered as a contrast subgraph if it satisfies the following two conditions: (1) $C$ is not subgraph isomorphic to subgraphs of any element in $G_a$. (2) One or more elements of $G_b$ are subgraph isomorphic to $C$ \cite{dong2012contrast}. In addition, Ting \textit{et al.} \cite{ting2006mining} proposed the concept of a minimal contrast subgraph to reduce the number of contrast patterns to be mined. When all subgraphs of the contrast subgraph $C$ are not contrast subgraphs, $C$ is the minimum contrast subgraph.
\end{definition}

Contrast subgraphs can distinguish between different graph datasets by revealing structural differences between them. Ting \textit{et al.} \cite{ting2006mining} proposed a method to find the minimal contrast subgraph, which is a graph pattern that appears in one graph but not in the other graph, and all of its proper subgraphs are either shared by or not contained in the two graphs. Wang \textit{et al.} \cite{wang2008spatial} and Gionis \textit{et al.} \cite{gionis2016bump} studied how to find the anomalous subgraphs that contrast others in one graph. Yang \textit{et al.} \cite{yang2018mining} adopted subgraph density as the measure for mining contrast subgraphs. Jiao \textit{et al.} \cite{jiao2020sub} proposed a scalable self-supervised graph representation via subgraph contrast, SUBG-CON.

\begin{definition}[Converging Patterns \cite{li2013efficient}] 
	\rm Suppose there are two datasets: $D_1$ belongs to class $c_1$ and $D_2$ belongs to class $c_2$. Then sup($X$, $D_1$) and sup($X$, $D_2$) denote the supports of itemset $X$ in $D_1$ and $D_2$, respectively. If an itemset $X$ satisfies the following conditions: \textit{CPs} = $\{X|sup(X,D_1)$ $\leq k$ $\times$ $sup(X,D_2)^\delta $, $sup(X,D_2)>\Theta\}$, where $k$ is the contrast coefficient, $\delta$ is the converging exponent, and $\Theta$ is the threshold of the minimal support in $D_2$, then the itemset $X$ is said to be the converging pattern. Converging pattern refers to the itemsets whose supports contrast sharply from the minority class to the majority one \cite{li2013efficient}. It can significantly differentiate itemsets in the class imbalanced data. Li \textit{et al.} \cite{li2013efficient} first introduced the concept of converging pattern, and proposed the ConvergMiner algorithm to discover converging patterns.
\end{definition}

\begin{definition}[Conditional Contrast Patterns, CCPs \cite{he2019mining,li2017pattern}] 
	\rm CCPs are a subset of contrast patterns (CPs). CCPs draw on the idea of conditional association to define a new concept, namely conditional contrast, which is based on the existence of sub patterns. Conditional contrast aims to filter the set of contrast patterns. Li \textit{et al.} \cite{li2017pattern} proposed an algorithm based on tree search for mining CCPs. Here, CCPs are a subset of traditional contrast patterns in one kind of dataset, and ``conditional" means a property of these patterns in another kind of dataset. He \textit{et al.} \cite{he2019mining} proposed conditional discriminative sequential pattern to solve the subset-induced redundancy issue. Here, ``conditional" means the redundancy of each pattern.
\end{definition}

\begin{definition}[Minimal Distinguishing Subsequence, MDS \cite{iqbal2017activity,ji2007mining}] 
	\rm A key feature of MDS is that items do not need to appear consecutively, which means that gaps can exist between items. Suppose there are two sequences: $s$ = $e_1$, $e_2$, $e_3$, ..., $e_n$ and $s'$ = $e'_1$, $e'_2$, $e'_3$, ..., $e'_m$. $s'$ is a subsequence of $s$. The occurrence $o_s$ of $s'$ in $s$ is a sequence of indices \{$i_1$, $i_2$, $i_3$,..., $i_m$\} if 1 $\le$ $i_t$ $\le$ $n$, 1 $\le$ $t$ $\le$ $m$, $e'_t$ = $e_{i_t}$, and $i_t$ < $i_t+1$. The gap constraint is denoted by $g$. The occurrence $o_s$ fulfills the g-gap constraint if $i_{k+1}$ - $i_k$ $\le$ g + 1 for each 1 $\le$ $k$ $\le$ $m$-1. Subsequence $s'$ satisfies the g-gap constraint when at least one occurrence of $s'$ satisfies the g-gap constraint.
\end{definition}

Suppose there are two sequences $s_1$ which belongs to class $c_1$ and $s_2$ which belongs to class $c_2$. Then sup($p$, $g$, $s_1$) and sup($p$, $g$, $s_2$) denote the supports of itemset $p$ in $s_1$ and $s_2$, respectively, where $g$ is a maximum gap. If an itemset $X$ satisfies the following conditions: sup($p$, $g$, $s_1$) $\ge$ $\delta$ and sup($p$, $g$, $s_2$) $\le$ $\alpha$. If no subsequence of $p$ satisfies previous two conditions, then the itemset $X$ is said to be MDS with $g$-gap constraint. MDS is a type of contrast pattern proposed by Ji \textit{et al.} \cite{ji2007mining}. The algorithm for mining MDS can be divided into three steps: generating candidate subsequences; calculating the support of subsequences; and removing subsequences \cite{iqbal2017activity}.

\begin{definition}[Emerging Patterns, EPs] 
	\rm EP is a subset of CP. Table \ref{EP_Types} summarizes the common types of emerging patterns. Figure \ref{eptype} \cite{garcia2018overview,loyola2020review} illustrates the relationships between these patterns. Emerging patterns were proposed by Dong \textit{et al.} in 1999 \cite{dong1999efficient} and are defined as itemsets with significantly increased support from one dataset to another. Trends that appear in the timestamp database can be captured by EPs. EPs can also find useful contrasts between data classes.

	\begin{table}[h]
		\caption{Types of emerging patterns}
		\label{EP_Types}
		\begin{tabular}{lp{5cm}p{8.3cm}}
			\toprule   
			\textbf{Abbreviation} &\textbf{Name} &\textbf{Description}\\
			\midrule 
			EP & Emerging pattern &  A pattern  $p$ is an emerging pattern if its support ratio in dataset $D^+$ and $D^-$ is not less than a given threshold value $\rho$, which can be expressed as EP = \{$p|GR(p)$ = $\frac{sup(p,D^-)}{sup(p,D^+)}  \geq \rho$\} \cite{dong1999efficient}.  \\  
			
			FEP & Fuzzy emerging pattern & FEP alleviates the disadvantage that general EP depends on clear numerical characteristic boundaries. FEPs are patterns generated by fuzzy selectors [Attribute $\in$ FuzzySet] \cite{garcia2011fuzzy}. \\ 
			
			JEP  & Jumping emerging pattern & Suppose a pattern  $p$ satisfies the following conditions:  $sup(p,D^+)$ = 0, $GR(p)$ = $ \frac{sup(p,D^-)}{sup(p,D^+)}$ = $\infty$ , then  $p$ is said to be the JEP \cite{dong1999discovering}. In short, JEPs are EPs that contain only the items in dataset $D^-$.
			\\ 
			
			Minimal EP & Minimal emerging pattern & An emerging pattern  $p$ is a minimal EP if none of its sub-patterns is an EP \cite{ventura2018supervised}. \\

			Maximal EP & Maximal emerging pattern &  If none of the sup-patterns of the emerging pattern  $p$ is an EP, then  $p$ is a maximal EP \cite{wang2004exploiting}.\\
			
			SJEP & Strong jumping emerging pattern & SJEP is also called Essential jumping emerging pattern(EJEP) in  \cite{fan2002efficient}. For a JEP  $p$, if there is no sup-pattern of  $p$ that is a JEP, then $p$ is a SJEP \cite{fan2006fast}.\\ 
			
			NEP & Noise-tolerant emerging pattern &  NEP is also called constrained emerging pattern(CEP). Suppose there are two threshold $\lambda_1$ and  $\lambda_2$. An emerging pattern  $p$ satisfies the following conditions: $\lambda_2
			\gg \lambda_1 $, $sup(p,D^+)$ $\leq \lambda_1$ and $sup(p,D^-)$ $\geq \lambda_2 $, then  $p$ is said to be a NEP \cite{fan2006fast}. \\ 
			
			Chi-EP & Chi emerging pattern &  Chi-EP is similar in concept to NEP, with the difference that Chi-EP use a $\chi^2$ test to measure differences in distributional significance \cite{fan2004noise}. \\ 
			
			BEP & Balanced emerging pattern & Suppose a pattern $p$ satisfies the following conditions:  $\frac{ae^{k*sup(p,D^+)}} {sup(p,D^-)} \geq \delta    \ and \sup(p,D^+) > \theta $, where $a$ is the correlative correction parameter, $\frac{e^{k*sup(p,D^+)}} {sup(p,D^-)}$ is the actual degree of growth of  $p$,	$k$ is the balance factor,	$\delta$ is the minimum support threshold,	$\theta$ is the minimum contrast coefficient, then  $p$ is said to be the BEP \cite{chen2016finding}.\\ 
			
			JEPN & Jumping emerging patterns with negation & A negated item indicates that the item is not included in the transaction. A negated value indicates that the value does not exist in the example. JEPNs allow JEPs to contain negated items \cite{terlecki2007jumping}. JEPNs can be defined as: JEPNs = \{$P | GR(p)$ = $\frac{sup(p,D^-)}{sup(p,D^+)}$ = $\infty$ \}, where $ P$ = \{$x_1$, $x_2$, ..., $x_n$ | $x_i \in \mathcal{P}$\}, $\mathcal{P}$ = $ \{x \subseteq \mathcal{I} \bigcup \overline{ \mathcal{I}} | $${\forall}$$_{i \in \mathscr{I}} i \in x \rightarrow \overline{i} \notin x \}$, $\mathcal{I}$ is the collection of existing selectors, $\overline{\mathcal{I}} = \{\overline{i}\}_{i \in \mathcal{I}} $ is the collection of selectors with negated values, $\mathscr{I}$ is an item space \cite{garcia2018overview, terlecki2007jumping}. \\
			
			\bottomrule  
		\end{tabular}
	\end{table}
	
	\begin{figure}[h]
		\centering
		\includegraphics[height=6cm]{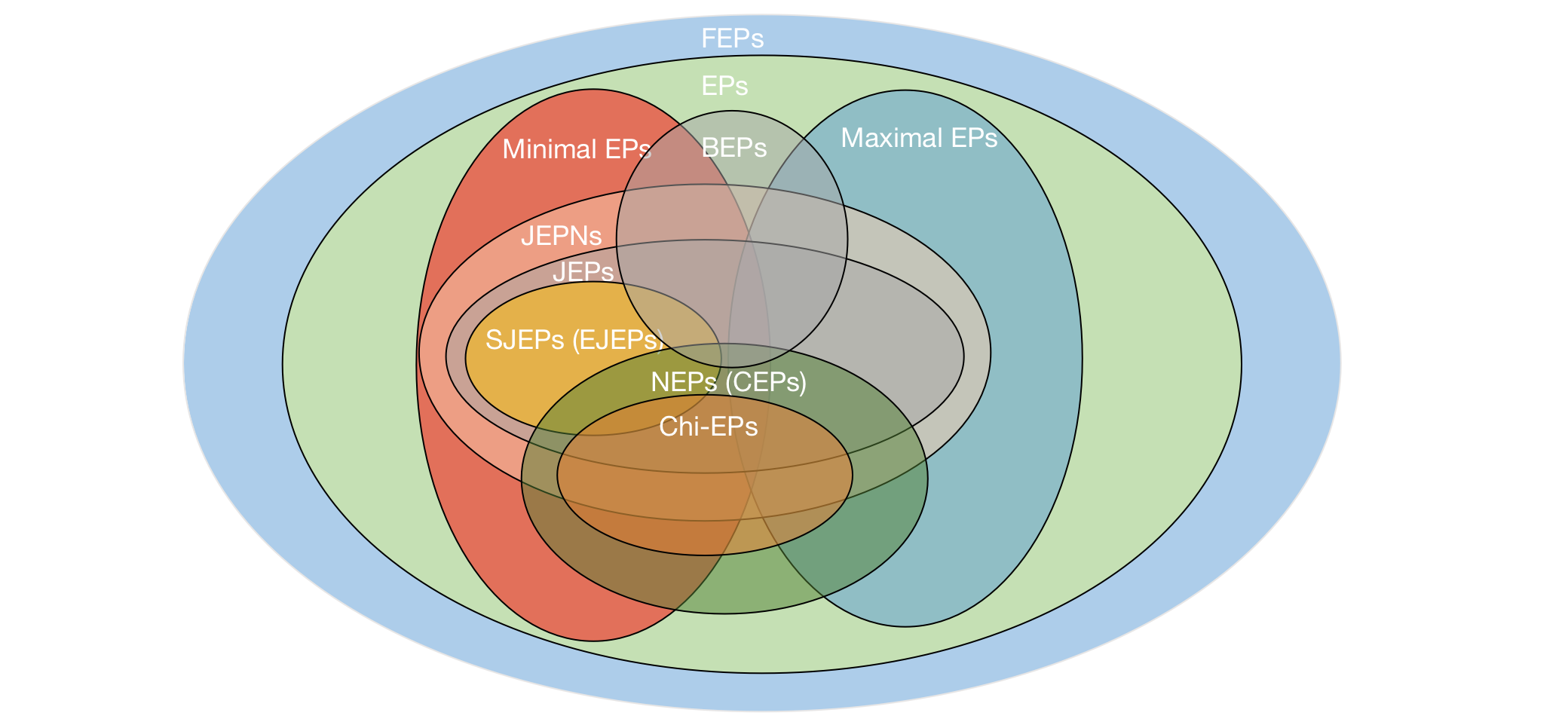}
		\caption{Relationships between different types of emerging patterns.} \label{eptype}
	\end{figure}	
	
\end{definition}

\subsection{Mining Strategy for Contrast Pattern Mining}

The mining strategies of existing contrast pattern mining algorithms can be divided into two categories. 
The first category is the mining strategy based on user-specified thresholds, which narrow the search space and filter out unwanted patterns by setting thresholds for some statistical measures, with the goal of finding all patterns that satisfy a given threshold \cite{zaiane2007finding}. 
The second type of mining strategy is based on statistical significance ranking, known as top-$k$ based contrast pattern mining algorithms, where the goal is to discover the top-$k$ patterns with high statistical significance of the method.

Most contrast pattern mining algorithms based on user-specified thresholds can be divided into two steps: first generating candidate patterns and then checking whether the contrast of the candidate patterns is greater than the threshold. The ConSGapMiner algorithm uses two support thresholds to discover the contrast patterns, meaning that all subsequences that satisfy the threshold and minimization conditions are put into the contrast pattern result set \cite{ji2007mining}. Most threshold-based contrast pattern mining algorithms set thresholds based on the contrast rate and support of the pattern. Different algorithms calculate the contrast rate in different ways. For example, the eCSP algorithm calculates the contrast rate of a pattern using the growth rate, and then uses the minimum contrast $\textit{min}_{cr}$ and minimum support $\textit{min}_{sup}$ thresholds to discover useful patterns \cite{zheng2016effective}.

Top-$k$ based contrast pattern mining algorithm has the advantage over the first type of algorithm. It can avoid the need for the user to set inappropriate thresholds, and the user only needs to set the number of patterns that they want to discover. The method is widely used in the fields of sequence patterns, association rules, sequence rules, and so on. In 2007, Osmar \textit{et al.} proposed the definition of top-n emerging sequence and the TopES algorithm, which mines the contrast patterns that occur most frequently in two datasets \cite{zaiane2007finding}. Although the patterns mined by TopES are only the significant sequences in the datasets and not the difference patterns, TopES is still important as an early top-$k$ mining algorithm. The kDSP- Miner algorithm, which was proposed in 2015, uses the top-$k$ method instead of support thresholds to discover contrasting sequence patterns \cite{yang2015mining}. The paper used top-$k$ to discover the $k$ most contrasting sequence patterns and build a classifier to achieve an effective method for diagnostic gene discovery \cite{zhao2015finding}. Previous studies have not considered the problem of mining distinguished sequential patterns in event sequences. Based on this problem, Duan \textit{et al.} proposed the concept of distinguishing temporal event patterns (DTEP), and introduced an algorithm to discover the top-$k$ contrasting significant DTEPs \cite{duan2017mining}. The previously mentioned algorithm does not consider the gap constraint well. Adaptively adjusting the gap constraint is one of the advantages of the SCP-Miner algorithm \cite{wu2021top}. SCP-Miner can find top-$k$ adaptive contrast patterns from positive and negative sequences. Top-$k$ mining strategy can also be applied to itemsets. For example, the paper addresses the proposed problem of how to use sampling methods to discover itemsets for discriminative sequences that are independent of the selected quality measure, and introduces two algorithms, called SeqScout and MCTSExtent, to extract the top-$k$ best non-redundant discriminative patterns and discover relevant subgroups in itemset-based sequences \cite{mathonat2021anytime}.

\textbf{Discussions}: The first type of algorithm is suitable for users who want to filter out all contrast patterns that satisfy the conditions, and the second type of algorithm is suitable for users who want a given number of contrast patterns with high statistical significance. Most of the current mining algorithms based on top-$k$ sorting strategies are focused on sequential contrast patterns, while other types of contrast patterns are less studied, such as episodes, subgraphs, events, and streams. In addition, most mining algorithms do not consider constraints, such as density constraints, length constraints, minimization pattern constraints, etc. Constraints can reduce the number of found patterns and lead to more interesting patterns.

\subsection{Statistical Measures of Discriminative Ability Assessment }

When using contrast patterns, an important task is to assess their discriminative ability \cite{bailey2016statistical}. Discriminative ability can be used as a way to assess the feature recognition ability of contrast patterns, or as a way to determine whether a pattern is a contrasting pattern. In this section, we summarize a series of common evaluation metrics used to assess the discriminative ability of individual patterns or groups of patterns. The common methods used to assess the discriminative ability of contrast patterns in two datasets can be generalized to multiple datasets. To facilitate the illustration of comparative assessment metrics, this section only describes how to apply the assessment metrics in the two datasets \cite{liu2015discriminative}.

First, we describe some terms that will be used to explain the comparative assessment indicators. Suppose there is a database $D$ with a set of $n$ transaction, $D$ = \{($x_i$, $y_i$)\}$_{i=1}^n$, where ($x_i$, $y_i$) is a transaction, $I$ = \{$i_1$, $i_2$, ..., $i_I$\}, $x_i$ $\subseteq$ $I$ is a itemset, and $y_i$ $\in$ \{+, -\} is the class label for $x_i$. We denote the dataset with all instances of class label + as $D$ $^+$ and the dataset with all instances of class label - as $D$$^-$, where $D$$^+$ $\cup$ $D$$^-$ = $D$. A pattern $p$ = \{$i_1$, $i_2$, ..., $i_k$\} $\subseteq$ $I$. Table \ref{number} \cite{liu2015discriminative} shows the number of pattern $p$ in databases $D$$^+$ and $D$$^-$, where $n_{11}$ denotes the number of patterns $p$ in $D^+$ and
|$D$|  denotes  the number of transactions for D.
The support of $p$ in $D$, $D^+$, $D^-$ respectively is denoted as:

\begin{equation}
$$
sup(p,D) = \frac{n_{1}}{|D|}, 
sup(p,D^+) = \frac{n_{11}}{|D^+|}, 
sup(p,D^-) = \frac{n_{12}}{|D^-|}.
$$
\end{equation}

\begin{table}[H]
	\centering
	\caption{A table of the number of pattern $p$}
	\label{number}
	\begin{tabular}{llll}  
		\toprule   
		& $D$$^+$ & $D$$^-$  & Sum\\  
		\midrule  
		$p$ & $n_{11}$ & $n_{12}$ & $n_{1}$  \\
		$\overline{p}$ & $n_{21}$ & $n_{22}$ & $n_{2}$  \\
		Sum & |$D$$^+$| & |$D$$^-$|  & |$D$| \\
		\bottomrule  
	\end{tabular}
\end{table}

\begin{table}[H]
	\caption{Summary of discriminative ability assessment approaches}
	\label{contrasting_assessment_approaches}
	\begin{tabular}{llll}  
		\toprule   
		Name & Measures \\  
		\midrule  
		Growth Rate & GR($p$,$D^+$,$D^-$) =  $\dfrac{sup(p,D^+)}{sup(p,D^-)}$\\
		
		Support Different & DiffSup($p$,$D^+$,$D^-$)= $| sup(p,D^+) - sup(p,D^-)| $ \\
		
		Unusualness & WRAcc($p$,$D^+$,$D^-$) = $\dfrac{n_1}{|D|}$($\dfrac{n_{11}}{n_1}$ - $\dfrac{|D^+|}{|D|}$) \\
		
		Generalization Quotient & $q_g$($p$,$D^+$,$D^-$) = $\dfrac{n_{11}}{n_{12}+g}$	\\
		
		OddsRatio & OddsRatio($p$,$D^+$,$D^-$) = $\dfrac{sup(p,D^+) / (1 - sup(p,D^+) )}{sup(p,D^-) / (1 - sup(p,D^-)}$ \\
		
		Gain & Gain($p$,$D^+$,$D^-$) = sup(p,D$^+$) $\times$ (log$\dfrac{sup(p,D^+)}{sup(p,D^+ \cup D^-) }$ - log$\dfrac{|D^+|}{|D^+| \cup |D^-|}$)
		\\
	    SupMaxK & SupMaxK($p$,$D^+$,$D^-$) = $ sup(p,D^+)$ - $max_{\beta\subseteq p}(sup(\beta,D^-))$
	    \\
	    MutualInformation & MI($p$,$D^+$,$D^-$) = $\sum_{i=1}^{i=2}$ $\sum_{j=1}^{j=2}$ $\dfrac{n_{ij}}{|D|}log\dfrac{n_{ij}/|D|}{n_i|D_j|/|D|^2}$
	    \\
	    Chi-squared & $\chi^2$ = $\sum_{i=1}^{i=2}$ $\sum_{j=1}^{j=2}$ $\dfrac{(n_{ij}-E_{ij})^2}{E_{ij}}$,$E_{ij} = \dfrac{\sum_{k=1}^{k=2}n_{ik} \sum_{k=1}^{k=2}n_{kj}}{|D|}$
	    \\
	    p - value & p - value = $\sum_{i=0}^{min(n_12,n_21)}\dfrac{t_1!t_2!|D^+|!|D^-|!}{|D|!(n_{11}+i)!(n_{12}-i)!(n_{21}-i)!(n_{22}+i)!}$
	    \\
		True Positive Rate &  TPR = $\dfrac{n_{11}}{n_1}$ \\
		False Positive Rate & FPR = $\dfrac{n_{21}}{n_2}$ \\
		Strength & Strength($p$) = $\dfrac{sup(p,D^+)^2}{sup(p,D^+) + sup(p,D^-)}$
		\\
		\bottomrule  
	\end{tabular}
\end{table}

Table \ref{contrasting_assessment_approaches} \cite{garcia2018overview,garcia2021cellular,liu2015discriminative,lucas2017new} shows the popular comparative assessment indicators. These evaluation indicators are important for eliminating unimportant patterns and obtaining accurate knowledge. In contrast pattern mining, how to choose assessment metrics is a question worth thinking about. It is not reasonable to talk about the merits of these metrics in isolation from the problem itself, so before selecting evaluation indicators, researchers need not only clarity about the problem needs but also likely insight into the domain knowledge. The evaluation metrics in Table \ref{contrasting_assessment_approaches} are described as follows:

\begin{itemize}
	\item[-] Growth Rate (GR). GR is based on the calculation of support, which calculates the ratio of support between two different classes. Dong \textit{et al.}  \cite{dong1999efficient} first use GR as an metric to evaluate contrast. If GR of pattern p not less than the user-given threshold (gr), then p is considered to be a contrast pattern. 
	
	\item[-]  Support Different (DiffSup). DiffSup is support difference between positive class and negative \cite{bay1999detecting}.
	
	\item[-]  Unusualness (WRAcc). WRAcc is also known as Weighted Relative Accuracy. WRAcc measures the balance between pattern generality and confidence. a high value of WRAcc indicates a high balance between generality and confidence \cite{garcia2018overview}.
	
	\item[-]  Generalization quotient ($q_g$). WRAcc and $q_g$ are two popular methods to evaluate the comparative ability of subgroups. The $g$ in the formula for $q_g$ is a user-defined parameter \cite{liu2015discriminative,gamberger2002expert}.
	
	\item[-]  OddsRatio. OddsRatio is a biometric measure that are typically used to assess simple risk factors. 
	Linking OddsRatio to pattern mining algorithms was first proposed by Li \textit{et al.}  \cite{li2007strong}, making it possible to discover compound risk factors in large-scale datasets. However, OR alone cannot infer whether the probability of occurrence or relative probability of occurrence of pattern $p$ in two categories is high or low \cite{li2007strong}.
	
	\item[-]  Gain. Gain can be used to assess the discriminatory ability of a pattern and to measure information about pos classes and other classes including that pattern \cite{liu2015discriminative}.
	
	\item[-]  SupMaxK. SupMaxK can not only find very low-frequency contrast patterns from high density and high dimensional datasets but also be used to prune the search space \cite{liu2015discriminative}. In addition, according to the calculation formula of SupMaxK and DiffSup, the lower bound of DiffSup is SupMaxK.
	
	\item[-]  Mutual information (MI). MI is based on information theory \cite{fang2011characterizing}. MI can be used to evaluate the correlation between pattern distribution and expected pattern frequency of each type \cite{garcia2013comparing}. 
	
	\item[-]  Chi-squared ($\chi^2$).  The chi-square test is used to test the independence of the variables in the column table \cite{bay2001detecting}. $E_{ij}$ is the expected frequency count of cell ij in cell ij given independence of the row and column variables in the table \cite{azevedo2010rules, liu2015discriminative}.
	
	\item[-]  p-value. The p-value can be used to measure the inconsistency between the original hypothesis and the sample. A smaller p-value means a stronger inconsistency. If the p-value is less than a predetermined threshold (significance level), the original hypothesis is considered to be rejected \cite{1928Statistical, None1926The}.
	 
	\item[-]  True positive rate (TPR). TPR shows the percentage of pattern $p$ in the positive dataset. It unites the precision and generality of the class \cite{garcia2019big,garcia2018overview}. 
	
	\item[-]  False positive rate (FPR). FPR descriptions the proportion of pattern $\overline{p}$  in the neg dataset. FPR can be used to measure generality and precision. Normally, we minimize the FPR value of pattern \cite{garcia2019big,garcia2018overview}.
	
	\item[-]  Strength. Strength is expressed by calculating the support of pattern p in each of the two classes or by calculating GR(p) and $sup(p,D^+)$ \cite{ramamohanarao2007patterns}. The strength satisfies the following formula:
	\textit{Strength}($p$) = $\dfrac{sup(p, D^+)^2}{sup(p, D^+) + sup(p, D^-)}$ = $\dfrac{GR(p, D^+, D^-)}{GR(p, D^+, D^-)+1}*sup(p, D^+)$ \cite{ramamohanarao2007patterns}. 
	
	From the formula, we can see that Strength(p) =  $sup(p,D^+)$ when $GR(p, D^+, D^-)$ = $\infty$. 
\end{itemize}

\subsection{Applications of CPM}

A major research theme in CPM in recent years has been the application of CPM to real-life situations. Many relevant papers have been presented to solve real-world problems. Interesting knowledge can be gained by applying CPM algorithms to real-world datasets. The CPM algorithm has been applied to solve problems in several areas, as follows:

\begin{itemize}
	\item[-] \textit{Medical domains}. CPM is widely used in the medical field. R{\'\i}os-M{\'e}ndez \textit{et al.} \cite{rios2020discovering} applied the EPM algorithm to a dataset of medical opinions about the decreasing number of autopsies in hospitals, with the aim of discovering medical experience in terms of doctors' choice or refusal of autopsies. In addition, EPM can be used to predict myocardial ischemia, diagnose coronary artery disease \cite{piao2016emerging} and discover toxicological knowledge \cite{sherhod2013toxicological,sherhod2012automating,sherhod2014emerging}.
	
	\item[-] \textit{Education domain}s. Researchers are increasingly employing data mining techniques to solve difficulties in the field of education. CPM is one of the data mining techniques that can be used to investigate the differences and similarities in learning patterns of different groups of students \cite{kong2020analysis} and to explore the relationships between learning attitudes \cite{thanasuan2017emerging}. In addition, contrast mining can be used in conjunction with association rule mining in education, for example, to obtain a contrast-based rule mining model \cite{tian2016discovering}.
	
	\item[-] \textit{Other applications}. CPM is also used in other fields. In the security domain, CPM can be used to build one-class of anomaly detection \cite{dong2018oclep+} and malware detection \cite{hellal2016minimal}. CPM can also be applied to bioinformatics to predict gene expression patterns \cite{dong2009applications}. In the field of cheminformatics, EP is used as an emerging chemical pattern (ECP), such as using ECP to predict the activity of compounds \cite{namasivayam2013prediction}. CPM can also be utilized to solve actual photovoltaic-related issues \cite{garcia2016analysing}. As a final example, CPM is applied to commodity applications, such as using EP to identify product sales trends \cite{weng2018observation} for smart retail.
\end{itemize}

\section{Methods of Contrast Pattern Mining}

During the past decades, a significant number of contrast pattern mining algorithms have been proposed. According to the mining principles and available data structures, we classify these algorithms into the following categories: 1) border-based algorithms; 2) tree-based algorithms; 3) evolutionary fuzzy system-based algorithms; 4) decision tree-based algorithms; and 5) others.
 
\subsection{Border-based Algorithms}

Dong \textit{et al.} \cite{dong1999efficient} proposed emerging patterns (EPs) for the first time and used border-based method to discover emerging patterns. Itemsets with significant increases in support from one dataset to another are emerging patterns.  Borders are used to describe large collections of itemsets or other types of collections. The concept of border and the typical algorithms are described below: 

\begin{definition}[Border \cite{dong1999efficient, li2004deeps, ventura2018supervised}]
	\rm $\mathcal{L}$ is a set of minimal emerging patterns. $\mathcal{R}$ is a set of maximal emerging patterns. An ordered pair $<$$\mathcal{L}$, $\mathcal{R}$$>$  is  a border, when the following conditions are satisfied:
	
	\begin{itemize}
		\item Each element of $\mathcal{L}$ and $\mathcal{R}$ is an anti-chain collection of sets. An anti-chain is a collection whose elements are subsets, and no element contains another, that is, for an anti-chain $S$, ${\forall}$ $M, N$ $\in$ $S$: $M$ $\nsubseteq$ $N$, $N$ $\nsubseteq$ $M$.
		
		\item Each element of $\mathcal{L}$ is a subset of some element in $\mathcal{R}$ and each element of $\mathcal{R}$ is a superset of some element in $\mathcal{L}$, which defined as ${\forall}$ $M$ $\in$ $\mathcal{L}$, ${\exists}$ $N$ $\in$ $\mathcal{R}$ such that $M$ $\subseteq$ $N$, ${\forall}$ $N \in \mathcal{R}$, ${\exists}$ $M$ $\in$ $\mathcal{L}$ such that $N$ $\supseteq$ $M$.
	\end{itemize}

\end{definition}

$\bullet$ MBD-LLBORDER \cite{dong1999efficient}. This algorithm can efficiently discover all EPs that satisfy the constraint, where borders are defined as the interval-closed collections represented by the borders' element. Large borders are discovered using Max Miner or HORIZON Miner, and the discovered borders are used as input to the MBD-LLBORDER algorithm which uses the borders to find the EPs that can be used to build the classifier. Besides, this algorithm introduces a multiborder-differential method to eliminate the need to check too many candidates.

$\bullet$ JEPProducer \cite{li2000space}. The concept of JEP space consists of all JEPs about a given positive and negative dataset. An algorithm is proposed and the property of JEP space satisfies convexity. That is, the JEP space is bounded and can be expressed concisely by boundary elements. This algorithm can be used to maintain the JEP space incrementally when extensive changes are made to previously processed data.

$\bullet$ DeEPs \cite{li2004deeps}. It is a lazy classifier, based on instances, where an instance is defined as a set of attribute-value pairs. The method uses bounds to concisely represent the EPs, which has the advantage of avoiding the need to enumerate the entire set of instance subsets during the discovery process and all patterns during the enumeration output. The found boundary EPs have many uses, such as: providing differentiation knowledge for our understanding of the instance $T$, utilizing the frequency of the boundary EPs to predict the class label of the test instance, and classifying $T$ if it is a test instance.

\subsection{Tree-based Algorithms}

Algorithms based on tree structure usually scan the dataset only once, and the data information is stored in the tree. Trees that mine contrast patterns usually have nodes that need to store information from both datasets. Patterns are formed by the path from the root to the node. The mining results are obtained based on the tree traversal. Mining of contrast patterns using tree structure has the following advantages: (1) Advantage in data storage. Since the tree can compress the representation of the dataset by sharing common prefixes \cite{garcia2016analysing}, it takes less time to scan the database and allows more data to be stored in the main memory. (2) Advantages in mining patterns and high mining efficiency. The items can be sorted according to certain values (e.g., growth rate) when building the tree. This makes the items with high comparative power closer to the root and allows them to be mined faster \cite{garcia2018overview}. In addition, such algorithms can employ pruning strategies to discard uninteresting patterns.

Algorithms based on tree structures face the following challenges: (1) Constructing a suitable tree. Table \ref{Tree} summarizes the common tree structures for mining the comparison patterns. These tree structures have different characteristics. Depending on the problem requirements, choose the appropriate tree structure or construct a new tree structure. (2) Choosing pruning strategies. Algorithms based on tree structures can easily face the problem that the search space is too large, resulting in high computational cost. Based on this problem, many pruning strategies have been proposed, such as: pruning non-minimal patterns \cite{terlecki2008efficient}, generating only promising branches \cite{fan2003efficiently}, and pruning strategies based on a priori principles \cite{liu2014novel}. One of the deciding factors for choosing a pruning strategy is the structure of the tree. (3) Redundancy in mining results. Detecting the redundancy of pattern sets is an important problem in contrast pattern mining that has not yet been solved.

\begin{table}[h]
	\caption{Tree-based algorithms for contrast pattern mining}
	\label{Tree}
		\begin{tabular}{lp{5.5cm}p{5.5cm}p{1.5cm}}
		\toprule   
		\textbf{Structure}&\textbf{Description}&\textbf{Characteristics}&\textbf{Ref.} \\
		
		\midrule 
	P-tree  & P-tree  is an ordered extended prefix tree structure, where order refers to the supports-ratio of items in descending order \cite{fan2002efficient}.  Closer to the root are items with higher supports-ratio. &Pruning strategies such as Chi-square test, data class superposition, generating only promising branches and anti-monotonicity conditions can be used. & \cite{fan2002efficient,fan2003bayesian,fan2003efficiently,li2017pattern}	\\ \hline

    CP-tree  & CP-tree is similar to P-tree in that both are ordered multi-way tree structures, the difference is that CP-tree has no node links \cite{fan2006fast}. The number of items in each node of the CP-tree is changeable. & CP-trees can be used to discover non-monotonic patterns and mine SJEPs and top-$k$ minimal JEPs. Pruning strategies are based on the anti-monotonicity or a-priori principle \cite{liu2014novel}. & \cite{fan2006fast,terlecki2008efficient} \\\hline

   DGCP-tree & The DGCP tree is an ordered tree structure that is used to store growth patterns and arrays of path codes from bit-string compression trees. Each node registers two arrays of path codes \cite{liu2014novel}. &The difference between DGCP-tree and CP-tree is that there is no need to scan the dataset to build a DGCP-tree, and there is no need to merge any nodes when accessing the nodes of a DGCP-tree \cite{liu2014novel}.& \cite{liu2014novel} \\\hline
   
   TDCSP-tree & It is a contrast sequential pattern suffix tree with information on time distribution \cite{wu2019mining}. This method requires only one scan of the database and stores data in a tree.  & The difference between a TDCSP-tree and a suffix tree is that the root is not empty. The root of a TDCSP-tree is a node containing a set of items \cite{wu2019mining}. &  \cite{wu2019mining}\\\hline
 
 CSP-tree  & The nodes of the CSP-tree need to record the prefix frequency of each class. The CSP-tree structure does not require the generation of candidate sequences \cite{zheng2016effective}. & The CSP tree can be pruned using Apriori-like pruning and some support-based pruning approaches, such as  heuristic cardinality pruning approaches \cite{zheng2016effective}. & \cite{zheng2016effective}\\\hline
 
 Nettree & Nettree is a tree-like data structure where the number of roots can be greater than one and there may be multiple paths from a node to its ancestors \cite{wu2019mining}. &The same node label might appear several times at different levels in Nettree, which effectively determines Whether or not a sequence of characters can be reused \cite{wu2019mining}.&  \cite{wu2019mining}\\\hline

 SLD-tree & It is a kind of subsequence position distribution suffix-tree, which can analyze the position distribution difference of different subsequences adaptively  \cite{li2019mining}. & In contrast to the null root of a suffix-tree, the root of the SLD-tree is also a node that holds a set of objects. Each item has six fields \cite{li2019mining}. &  \cite{li2019mining}\\\hline

 Prefix-leaf tree & A prefix leaf tree is a type of prefix tree in which each internal node holds the symbols of its leaf node offspring, regardless of whether its parent is the root node or has numerous children.& All candidate patterns can be organized using the prefix-leaf tree. Depth-first traversal is used to find the collection of candidate patterns that match \cite{wang2014efficient}.&  \cite{wang2014efficient}\\\hline

 T*-tree  & The T*-tree (transaction tree) structure is divided into trunk and branch. By gain ratio, the attributes are listed in descending order, with the top 10$\%$ attributes as trunk nodes and the rest attributes as branch nodes \cite{li2013efficient}. & A T*-tree is an index structure that asks how many transactions match a particular pattern, where transactions being represented  as spatial objects wrapped by Minimal bounding boxs (MBBs) \cite{li2013efficient}.&  \cite{li2013efficient}\\\hline

 FP-tree, TFP-tree & The node of FP Tree records the project frequency of each class \cite{bailey2002fast,li2007mining}. TFP-tree means generating a T-tree from a P-tree using a similar priori approach, where T-tree is a compressed set enumeration tree \cite{piao2011enumeration}.  & FP Tree is an early structure used to discover JEPs \cite{bailey2002fast,li2007mining}. Compared to FP-tree, TFP-tree has shorter generation time and is more efficient in memory management \cite{piao2011enumeration}.  &  \cite{bailey2002fast,li2007mining,piao2011enumeration} \\
         
		\bottomrule  
	\end{tabular}
\end{table}

\begin{table}[h]
	\caption{Decision tree-based algorithms for contrast pattern mining}
	\label{DecisionTree}
	\begin{tabular}{lp{4cm}p{4cm}p{4cm}p{0.8cm}}
		\toprule   
		\textbf{Name} &\textbf{Description}&\textbf{Pros.}&\textbf{Cons.}&\textbf{Year}\\
		\midrule 
		
		LCMine \cite{garcia2010lcmine}
		& This method induces a series of decision trees and filtering operations to find high-quality discriminative attributes.
		& Discriminant rules can be efficiently found on training sample sets with incomplete and mixed data.
		& This method requires post-filtering of the schema, which is time-consuming.
		& 2010\\  \hline
		
		CEPM \cite{garcia2010new}
		& The improved version of LCMine uses a new weighting scheme to mine various patterns without global discretization.
		& CEPM is faster and more accurate than LCMINE. CEPM does not require schema filtering post-processing.
		& The operation of estimating the minimum support threshold is inefficient. & 2010\\  \hline
		
		EPRFm \cite{wang2010building}
		& The algorithm generates decision trees in the same way as random forest and mines decision rules for each data class.
		& EPRFm can be used for video database and get good classification results. 
		& For small data, the classification result of random forest is not good. & 2010\\ \hline
		
		FEPM \cite{garcia2011fuzzy}
		& The algorithm uses a set of fuzzy decision trees to extract fuzzy emerging patterns.
		& It first introduces the concept of fuzzy emerging pattern (FEP) and the algorithm of mining FEP. 
		& The same pattern will appear on different trees, resulting in time consuming. & 2010\\  \hline
		
		DBF \cite{M2015Finding}
		& The core idea of DBF is to remove certain attributes that appear in many patterns by removing features in the induced decision tree.
		& The situation of creating too many repeating patterns has been improved.
		& Deleting a feature directly causes other relationships for that feature to be deleted along with it, thus losing many of the resulting patterns.
		& 2015\\ \hline
		
		DBP \cite{M2015Finding}
		& From deleting features to deleting attributes, it does not allow the attributes of root nodes to exist in other induced decision trees.
		& DBP maintains the integrity of research results to a certain extent.
		& There may be attributes that have the same properties as the best attributes, making DBP less useful.
		& 2015\\\hline
		
		DBPL \cite{M2015Finding}
		& Remove the best attribute by level, allowing a given attribute to appear in a sublevel.
		& To a certain extent, it solves the inaccuracy of results caused by DBP's strict deletion.
		& Determining attribute levels is tedious and changeable.
		& 2015\\\hline
		
		RFM \cite{M2015Finding}
		& Random forest is one of the best methods to discover CP, which is most suitable for classifying data.
		& The classifier based on random forest has the advantages of low cost and high accuracy.
		& Due to the complexity of the algorithm, it takes more time to train and costs more than other algorithms. & 2015\\ \hline
		
		Bagging \cite{M2015Finding}
		& One of the best ways to discover CP and performs best when used with numerical data.
		& For any support threshold, Bagging's abstinence rate is low and accurate.
		& The algorithm requires more time to train than other algorithms.
		& 2015\\ \hline
		
		PBC4cip \cite{2017PBC4cip}
		& A strategy is proposed to solve the class imbalance problem. 
		& In class imbalance problem, PBC4cip can get good classification results.
		& Due to the limitations of the Hellinger distance, PBC4cip can only solve two kinds of problems.
		& 2017 \\ \hline
		
		MHRFm \cite{canete2019classification}
		& MHRFm uses Multivariate decision trees to deal with multiple types of problems.
		& MHRFm removes the restriction that HRFm handles only two types of problems.
		& MHRFm's filtering operation takes a long time to run.
		& 2019 \\ 
		\bottomrule  
	\end{tabular}
\end{table}

\subsection{Evolutionary Fuzzy Systems-based Algorithms}

The evolutionary fuzzy system-based algorithms is a rule extraction algorithm with better descriptive power than the previous classes of algorithms \cite{garcia2018moea}. It combines the concepts of fuzzy logic systems and natural evolution. The concept of a fuzzy logic system is relative to a traditional logic system. Traditional logic systems have values of either 0 or 1, whereas in fuzzy logic, a value between 0 and 1 can exist. Fuzzy rules are closer to the knowledge representation of human reasoning. Fuzzy logic can improve the quality of mining patterns and the interpretability of results. The principle of natural evolution is that organisms with access to resources tend to have offspring in the future. Evolutionary algorithms allow us to find a good solution in a large search space. A big data approach for extracting fuzzy emerging patterns has a reasonable amount of time \cite{garcia2018moea}. In an evolutionary algorithm, starting from an initial set of randomly generated individuals, the next generation of individuals is derived, following the genetic pattern of living organisms. Individuals are then selected according to their fitness to improve the quality of the next generation population, and after several iterations, the optimal solution is gradually approached. The evolutionary algorithm is essentially a search and find method of optimization.

$\bullet$ EvAEP \cite{garcia2016analysing}. It is an evolutionary fuzzy system algorithm for the extraction of emerging patterns, applied to the problem of concentrated photovoltaic technology. EvAEP uses an evolutionary algorithm. It considers the solution as an individual, represented as a connection of pairs of variable values defined in a discrete domain. When the variables are defined in the continuous domain, EvAEP uses a fuzzy logic where the fuzzy set consists of linguistic labels. These labels are defined by uniform triangular forms. The algorithm uses an iterative rule learning model to find high quality solutions. In each iteration of the evolutionary process, the current best solution is passed on to the next generation, which acquires new individuals by applying genetic operators to the previous generation.

$\bullet$  EvAEFP-Spark \cite{garcia2017first}. This algorithm is an optimization of the EvAEP algorithm. It applies the MapReduce paradigm, allowing the EvAEFP Spark algorithm to quickly discover EPs with highly descriptive features on large datasets. The main idea of EvAEFP Spark is to apply the MapReduce framework to modify the way in which individuals are evaluated during evolution, thus reducing the complexity when evaluating groups or sets of individuals \cite{ventura2018supervised}. Map and reduce are the two phases of MapReduce. In the map phase, the dataset is decomposed and each subset is analyzed in a different mapper for all the individuals in the population and an obfuscation matrix is obtained for each population \cite{carmona2018unifying}. In the reduce phase, all the obfuscation matrices are accumulated into a generic matrix and the fitness of each individual is calculated.

$\bullet$ MOEA-EFEP \cite{garcia2018moea}. It is a multi-objective evolutionary algorithm for the extraction of fuzzy emerging patterns. MOEA-EFEP has some commonalities with the EvAEP algorithm, for example, the encoding method is ``chromosome = rule" \cite{loyola2020review}. The main difference between them is that EvAEP is a multi-objective evolutionary algorithm. The main difference is that EvAEP is a single-objective based evolutionary algorithm, whereas MOEA-EFEP is multi-objective. The algorithm uses a cooperative-competitive model for elite groups, allowing simultaneous optimization of multiple quality measures, where individuals cooperate by maximising the average unusualness of the group and compete to obtain the optimal solution to the problem following a token competition procedure.

$\bullet$ E2PAMEA \cite{garcia2020e2pamea}. It is a cooperative competition multi-objective evolutionary fuzzy system, which adopts cooperative competition schema and token competition program to improve the reliability of results. The algorithm uses an evaluation process called Bit-LUT to improve the effectiveness of the method. Bit-LUT is a MapReduce method based on precise methods. Bit-LUT is faster and requires less physical memory than previous evaluation methods. E2PAMEA algorithm can be used in big data environment to efficiently extract high-quality EPs. Compared with the EPM evolutionary algorithm of big data proposed in the past, E2PAMEA algorithm has higher applicability and the results of mining are more reliable and universal.

$\bullet$ CE3P-MDS \cite{garcia2021cellular}. It is used to extract EPs in massive data streams from various sources through Apache Kafka streaming media platform and Apache Spark streaming media library. This algorithm combines the ideas of fuzzy logic and cell-based multi-objective evolutionary algorithm, and proposes a coverage-ratio-based approach, which is used to reduce the redundancy of patterns and and is also used as a reinitialization process. The algorithm uses an informed strategy, which is used to process data, and a reinitialization and filtering mechanism, which is used to reduce redundancy and low reliability patterns. Experiments showed that the algorithm is very fast in data processing, and it takes about 4 seconds to process more than 750,000 instances. In addition, massive, heterogeneous and high-speed data streams can be processed by the algorithm.

\subsection{Decision Tree-based Algorithms} 

A decision tree is an algorithm that can be used to solve classification problems. It is also a kind of supervised learning. This process is called supervised learning, which first builds a model from a bunch of samples of known categories, each with a set of attributes and a classification result, and then uses the model to predict the test sample set. A decision tree is a tree structure summarized from training data, in which each internal node represents a judgment on an attribute, each leaf node represents a classification result, and each branch represents the output of a judgment result. The general steps of the contrast pattern mining algorithm based on decision trees can be divided into the following two points: (1) Summarize different decision trees with training datasets and strategies to create diverse decision trees. (2) Extract contrast patterns from each induced decision tree \cite{garcia2014survey}. The time complexity of mining contrast patterns from all induced decision trees is $O(Tmnlog_2(n))$, where $T$ represents the number of established induced decision trees, $m$ is the number of features in the training dataset, and $n$ is the number of objects in the training dataset \cite{loyola2020review}.

Table \ref{DecisionTree} \cite{garcia2014survey,garcia2018overview,loyola2020review} summarizes the decision tree-based algorithms and lists the differences between them. Compared with other types of algorithms, the advantages of the CPM algorithms based on decision trees are as follows:

\begin{itemize}
	\item It is easy to understand, extract rules, and explain in a decision tree. And decision trees can be visually analyzed.
	
	\item High accuracy: The discovered classification rules have high accuracy and are easy to understand. The decision tree can clearly show which fields are more important; that is, understandable rules can be generated.
	
	\item Running speed is relatively fast.
\end{itemize}

The decision tree algorithms also have some disadvantages, such as:

\begin{itemize}
	\item  Over-fitting is prone to occur. A decision tree model tends to produce an overly complex model, which has poor generalization performance for data, which is called over-fitting. To put it simply, the decision tree describes the characteristics of training samples too accurately, which makes it impossible to analyze the new samples reasonably. The over-fitting phenomenon can be reduced by cutting the branches that affect the prediction accuracy. There are two common pruning strategies: pre-pruning and post-pruning. The pre-pruning strategies mainly restrict the full growth of decision trees by establishing some rules, while the post-pruning strategies prune after the full growth of decision trees.
	
	\item Different feature selection criteria will lead to different feature selection tendencies. For example, the information gain criterion has a preference for attributes with a large number of redeemable values (typically represented by the ID3 algorithm), while the gain rate criterion (CART) has a preference for attributes with a few redeemable values.
\end{itemize}

\subsection{Others}

We group the algorithms that do not belong to the previous categories into this section, such as the ZBDDs algorithm, which can handle higher dimensional data, and the GEP algorithm, which is based on the principle of natural selection.

$\bullet$ \textit{Zero-suppressed binary decision diagrams-based algorithms} \cite{loekito2006fast}. The Zero-suppressed binary decision diagram (ZBDD) is an excellent data structure for handling sparse data and high-dimensional datasets. Previous EP classifiers were incapable of handling datasets with more than 60 dimensions, while ZBDDs can solve this problem. ZBDDs use two reduction rules: the merging rule and the zero-suppression rules. The merge rule implies a shared equivalence subtree, and the zero-suppression rule means that nodes with real edges pointing to the rule allow high compression operations on Boolean formulas. The study \cite{loekito2006fast} uses disjunction and conjunction to generalize EP and names these patterns as disjunctive emerging patterns. Two algorithms are presented: mineEP and mineDEP. The mineEP algorithm takes ZBDD as the input parameter and uses a bottom-up approach to find the smallest EP, while the mineDEP algorithm uses a top-down approach to discover the largest disjunction EPs. According to the experimental results, the advantage of ZBDDs over pattern trees is that ZBDDs can handle higher-dimensional datasets and have shorter running times.

$\bullet$ \textit{Gene expression programming} \cite{gao2016mining}. Gene expression programming (GEP) selects individuals based on fitness, which is encoded as a fixed-length linear string and subsequently expressed as an expression tree. GEP can find the optimal solution by iterative evolution \cite{ferreira2001gene}. The GepDSP algorithm \cite{gao2016mining} was designed based on the GEP framework. Two types of sequences are used as input variables for GepDSP. The algorithm first generates a random number of individuals (considered as individuals of the first generation) to form the initial population. An evolutionary process is performed on each generation of individuals (including assessing the fitness of each individual, saving the $k$ individuals with the highest contrast score, selecting individuals, modifying the selected individuals, and then replicating them) until some conditions are met. The experiments validated the effectiveness and efficiency of the GepDSP algorithm on mining kDSPs with flexible gap constraints.

\section{Advanced Topics of CPM}

\subsection{Class Imbalance Problem in CPM}

In previous studies, CPM was mostly based on the assumption that the dataset distribution is basic and balanced. However, in the real world, it is a common phenomenon that datasets have class imbalance problems. The class imbalance problem refers to the problem of uneven distribution of objects into classes, which is manifested by the fact that the number of objects in some classes is much smaller than that in other classes \cite{loyola2019cost}. In some applications, people tend to be more interested in learning about rare or few classes of data in a dataset. For example, in credit card fraud detection, there are far fewer fraudulent transactions than legitimate transactions, but people are more concerned about fraudulent transactions than information about legitimate transactions. Many problems can arise when a classifier based on the assumption that the distribution of the dataset is balanced is applied to an unbalanced dataset. These classifiers tend to sacrifice the accuracy of the minority class in order to improve overall accuracy, ignoring the reality that in many actual applications, misclassification of the minority class samples is usually more expensive \cite{chen2016finding}. Many methods have been proposed to solve the class imbalance problem in the dataset. These methods can be classified into the following three categories: 1) data-level methods, 2) algorithm-level methods, and 3) cost-sensitive learning methods \cite{krawczyk2016learning,ng2021maximizing}.

\subsection{Fuzzy Contrast Pattern Mining}

EP-based classifiers usually contain a prior discretization operation, which is used to discretize individual attributes. This operation has the following two disadvantages: 1) The discretization process can easily result in information loss. 2) Numerical discretization usually defines explicit numerical feature boundaries, and pattern matching relationships lack flexibility. For example, the object (2,6) may match a pattern while (2.001,6) may not. To address these drawbacks, FEP was proposed by Garc{\'\i}a-Borroto \cite{garcia2011fuzzy}. FEP is a pattern that combines the concepts of fuzzy logic with emerging patterns to obtain a knowledge representation that is close to human reasoning. The advantage of FEP is to make patterns easier to read and interpret and to avoid information loss in the process of discretization \cite{garcia2019big}. So far, several research studies have been proposed. The MOEA-EFEP algorithm \cite{garcia2018moea} is based on the NSGA-II algorithm for mining EP. And a multi-objective evolutionary algorithm has been used to discover FEP. Researchers have also studied mining FEPs in different environments, such as data streams \cite{garcia2020fepds} and big data \cite{garcia2017first,garcia2020e2pamea,garcia2019big}.

\subsection{CPM in The Big Data Environment}

In the era of digital economy, Big Data is characterized by large volume, fast generation, diversity, accuracy, and value (the 5V's model) \cite{garcia2021cellular}. Mining contrast patterns in the Big Data environment is a meaningful research direction. Mining CP is a challenge due to the characteristics of the 5V's model of Big Data. Existing approaches to mining CPs from big data can be divided into the following two categories:

\begin{itemize}
	\item  Designing parallelized CPM algorithms, such as the DCP-Growth algorithm solves the problem by dividing the search space of CPs into small, independent units, where these units can be mined in a parallel manner, providing a scalable solution for mining large datasets \cite{savage2016distributed}.

	\item Developing CPM algorithms based on existing Big Data technologies (such as MapReduce \cite{2018MapReduce} and Spark \cite{zaharia2010spark}), such as the EvAEFP Spark algorithm and the BD-EFEP algorithm. The EvAEFP Spark algorithm uses the MapReduce paradigm, developed in Apache Spark, to efficiently analyze huge datasets \cite{garcia2017first,ventura2018supervised}. BD-EFEP uses the MapReduce method and evolutionary method to process large amounts of data efficiently \cite{garcia2019big}. 
\end{itemize}

\subsection{Data Visualizations}

Applying CPM to the data can obtain interesting information, but this information might be difficult to comprehend. Data visualization is used as a tool in data mining to better interpret data. Data visualization can not only help users to make sense of complex problems, but also discover other interesting phenomena by extracting and analyzing information. Combining research results with data visualization has been a popular strategy in recent years because it allows users to view data from a variety of angles. In 2015, a method for visualizing and predicting spatial-temporal events based on contrast patterns was proposed \cite{wang2015contrast}. In the following year, Nishiguchi \textit{et al.} proposed a visualization method based on path diagrams, called CRPD, which could be used to explain the relationships between CPs \cite{nishiguchi2016caecp}. The idea of visualization has been applied to practical problems. For example, Loyola-Gonz{\'a}lez \textit{et al.} present the first scientometrics study of world university rankings based on contrast patterns and a method for visualizing extracted patterns, with the aim of helping policy-makers to develop and evaluate strategies for ranking world universities \cite{loyola2020contrast}. In 2021, Neto  \textit{et al.} proposed a visual analysis method called VAX \cite{neto2021multivariate}.  VAX uses the descriptive capabilities of JEPs to analyze information. Matrix-like visual metaphors were used in this method in order to present JEPs.

\subsection{CPM with New Technologies}

The new generation of information technologies includes the Internet of Things (IoT) \cite{gubbi2013internet}, 5G \cite{rappaport2013millimeter}, cloud computing \cite{wang2010cloud}, blockchain \cite{swan2015blockchain}, and artificial intelligence \cite{2021Artificial}. Their combination with data mining plays an important role in promoting the development of the digital economy and creates infinite application value together. As a subfield of data mining, the combination of CPM and new technology is an advanced topic. 5G technology is distinguished by its high bandwidth, low latency, and high stability \cite{ahmad2019security}. 5G solves the problem of communication transmission efficiency, and the value will be reflected in data mining. Data mining uses efficient data transmission to analyze data. The combination of data mining and new technology has great application space and research value, such as the combination of data mining and the IoT \cite{sunhare2020internet,shadroo2018systematic}. In addition, the combination of data mining and various technologies has also been investigated by researchers. For example, a cloud platform based on cloud computing was applied to IoT data mining technology \cite{gaber2019internet,idhammad2018distributed}. The combined technology has many advantages, such as reducing the time of data transmission, improving the efficiency of data mining, and avoiding the failure of data storage.

\section{Challenges and Opportunities}

While research on CPM has been fruitful, there are still some key issues that need to be studied in depth. In this section, we discuss the open challenges and opportunities in this area.

\begin{itemize}
	
	\item \textit{Mining multivariate contrast patterns}. A univariate contrast pattern is a contrast pattern in which each term of the pattern involves only one feature. Most of the CPM studies are based on univariate contrast patterns, but they have limitations. In some datasets, the univariate contrast pattern-based classifier does not work well. Therefore, the concept of multivariate contrast patterns (MCP) has emerged, where the items of MCP can be either univariate or multivariate items \cite{canete2019classification}. MCPs have better classification performance than univariate contrast patterns, but they have the disadvantage of being more difficult to understand. At present, there is little research on MCP, but MCP is an interesting research direction. There are some challenging research opportunities for MCP: how to discover MCPs in different data environments; how to classify using fuzzy multivariate patterns; and how to design unsupervised classifiers based on MCPs.

    \item \textit{Open-source software}. Although CPM has been studied for more than 20 years, there is no well-known open-source platform that provides source code for studying CPM for researchers to exchange and learn. This is a problem that needs to be looked at. The lack of an open-source platform not only makes the exchange of source code difficult, but also hinders the further development of the field to some extent. This is because researchers often need to re-implement other researchers' algorithms in order to improve them or to compare the performance of new ones. Therefore, there is an urgent need to develop an open-source platform for contrast pattern mining.
	
	\item \textit{Datasets for CPM}. CPM research requires the support of open-source datasets. The lack of open-source datasets makes it difficult for researchers to evaluate the performance of algorithms, as too few experimental datasets are not conducive to discovering algorithm flaws and strengths, and experimental results lack credibility. One of the current research trends is to use algorithms to solve real-world problems directly, and the study of real-world problems often requires the support of relevant datasets, so the issue of whether these datasets are open source is very important. Some common open source datasets are listed below:
	
	\begin{itemize}
		\item[-] The UCI repository: https://archive.ics.uci.edu/ml/index.php
		\item[-] The FIMD repository: http://fimi.ua.ac.be/data
		\item[-] SPMF:  http://www.philippe-fournier-viger.com/spmf/index.php	
	\end{itemize}
	
	\item \textit{Integration with other concepts}. Adding concepts from other fields to CPM is a promising research direction. Researchers can consider combining contrast patterns (CPs) with other types of patterns. Take advantage of their features and strengths to solve practical problems effectively. Some researchers have already combined CPs and utility patterns to come up with the concept of the conditional contrast high-utility itemset. In addition, researchers can combine CPs with concepts from other fields to obtain new concepts. For example, FEP is the concept obtained by combining fuzzy logic with EP \cite{garcia2011fuzzy}. CPs combined with other concepts can also be used directly to solve certain problems. For example, EP combined with Random Forest is used to identify complex activities in smart homes \cite{malazi2018combining}.
	
	\item \textit{Dynamic data mining}. With the rapid development of data acquisition, storage, and transmission technologies, data is no longer static and can be updated dynamically. The timeliness of data is getting shorter and the scale of data is getting larger, and the use of traditional static data mining techniques to analyze the constantly generated information can no longer meet the realistic needs. In order to obtain the knowledge of interest in the constantly generated data streams, researchers have proposed many approaches to dynamically process real-time data, and related works include online data mining \cite{akhriza2015novel}, incremental pattern mining algorithms \cite{bailey2010efficient}, and streaming data mining \cite{garcia2020fepds}. There are still some research opportunities in dynamic data mining: how to develop CPM algorithms based on existing big data technologies such as MapReduce \cite{2018MapReduce}, Spark \cite{zaharia2010spark}, and Storm \cite{van2015dynamically}, and how to improve the efficiency of algorithms.
	
	\item \textit{Privacy \& Security}. Concern about data privacy and security has become a worldwide trend. Using private data to discover patterns will inevitably cause privacy problems, so how to discover patterns on the basis of protecting data privacy becomes an urgent problem to be solved. So far, researchers have proposed some methods to protect the privacy of frequent patterns \cite{2016DPcode, 2018Privacy, zhang2017efficient}. However, in the field of CPM, there is no research on the contrast pattern of privacy protection. Some application scenarios have privacy requirements. For example: marketing and healthcare. In these application scenarios, we need to protect the privacy of datasets and the security of mining results while mining contrast patterns. In order to expand the application scenarios of contrast patterns, the research of privacy-preserving contrast pattern mining should be given more attention.

	\item \textit{Contrastive learning}. Data mining is closely related to machine learning. On the one hand, machine learning provides the underlying technical support for data mining. On the other hand, machine learning needs a large amount of valid data for training. CPM belongs to data mining. Contrastive learning belongs to self-supervised learning in machine learning. In contrastive learning, positive and negative samples are compared in the feature space to learn the features of samples \cite{chuang2020debiased}. Contrastive learning focuses on learning the common characteristics between similar instances and distinguishing the differences between non-similar instances \cite{khosla2020supervised}. CPM compares positive and negative datasets to discover patterns of contrast. There are ideological similarities between CPM and contrastive learning. Contrastive learning can provide underlying technical support and thought inspiration for the CPM field.

\end{itemize}

\section{Conclusions}

Contrast pattern mining can be used to solve many real-world problems effectively, and other types of pattern mining methods actually use the concept of contrast. The combination of contrast pattern mining with other pattern mining methods is a very interesting research direction, and their combination may help discover more efficient solutions to old problems or discover and solve new problems. Therefore, contrast pattern mining is a very important research topic. However, there are few studies that summarize and classify contrastive pattern mining methods. In this study, we introduce the basic concepts of types of CPM, mining strategies, and assessment of comparative power. In addition, classical algorithms of CPM are classified and their advantages and disadvantages are introduced for analysis. Advanced and up-to-date topics related to CPM are also discussed. As an important area of data mining, although CPM research has been fruitful, there are still some key issues that need to be studied in-depth. Thus, open challenges and future directions of contrast pattern mining are presented at the end.

\section*{Acknowledgment}

This research was supported in part by the National Natural Science Foundation of China (Grant Nos. 62002136 and 62272196), Natural Science Foundation of Guangdong Province of China (Grant No. 2022A1515011861), Guangzhou Basic and Applied Basic Research Foundation (Grant No. 202102020277), and the Young Scholar Program of Pazhou Lab (Grant No. PZL2021KF0023).

\printcredits

\bibliographystyle{cas-model2-names}

\bibliography{SurveyCPM.bib}

\begin{thebibliography}{100}

\bibitem{cisco2021cisco}
U~Cisco.
\newblock Cisco annual internet report (2018--2023) white paper.
\newblock {\em Acessado EM}, 10(01), 2021.

\bibitem{chen1996data}
Ming-Syan Chen, Jiawei Han, and Philip~S Yu.
\newblock Data mining: an overview from a database perspective.
\newblock {\em IEEE Transactions on Knowledge and Data Engineering},
  8(6):866--883, 1996.

\bibitem{fournier2022pattern}
Philippe Fournier-Viger, Wensheng Gan, Youxi Wu, Mourad Nouioua, Wei Song, Tin
  Truong, and Hai Duong.
\newblock Pattern mining: Current challenges and opportunities.
\newblock In {\em International Conference on Database Systems for Advanced
  Applications}, pages 34--49. Springer, 2022.

\bibitem{gan2017data}
Wensheng Gan, Jerry Chun-Wei Lin, Han-Chieh Chao, and Justin Zhan.
\newblock Data mining in distributed environment: a survey.
\newblock {\em Wiley Interdisciplinary Reviews: Data Mining and Knowledge
  Discovery}, 7(6):e1216, 2017.

\bibitem{gan2019survey}
Wensheng Gan, Jerry Chun-Wei Lin, Philippe Fournier-Viger, Han-Chieh Chao, and
  Philip~S Yu.
\newblock A survey of parallel sequential pattern mining.
\newblock {\em ACM Transactions on Knowledge Discovery from Data}, 13(3):1--34,
  2019.

\bibitem{ventura2018supervised}
Sebasti{\'a}n Ventura, Jos{\'e}~Mar{\'\i}a Luna, et~al.
\newblock {\em Supervised descriptive pattern mining}.
\newblock Springer, 2018.

\bibitem{luna2019frequent}
Jos{\'e}~Mar{\'\i}a Luna, Philippe Fournier-Viger, and Sebasti{\'a}n Ventura.
\newblock Frequent itemset mining: A 25 years review.
\newblock {\em {Wiley Interdisciplinary Reviews: Data Mining and Knowledge
  Discovery}}, 9(6):e1329, 2019.

\bibitem{gan2021survey}
Wensheng Gan, Jerry Chun-Wei Lin, Philippe Fournier-Viger, Han-Chieh Chao,
  Vincent~S Tseng, and Philip~S Yu.
\newblock A survey of utility-oriented pattern mining.
\newblock {\em IEEE Transactions on Knowledge and Data Engineering},
  33(4):1306--1327, 2021.

\bibitem{dong2012contrast}
Guozhu Dong and James Bailey.
\newblock {\em Contrast data mining: concepts, algorithms, and applications}.
\newblock CRC Press, 2012.

\bibitem{alipourchavary2020improving}
Elaheh AlipourChavary, Sarah~M Erfani, and Christopher Leckie.
\newblock Improving scalability of contrast pattern mining for network traffic
  using closed patterns.
\newblock {\em arXiv preprint, arXiv:2011.14830}, 2020.

\bibitem{wu2019mining}
Youxi Wu, Yuehua Wang, Jingyu Liu, Ming Yu, Jing Liu, and Yan Li.
\newblock Mining distinguishing subsequence patterns with nonoverlapping
  condition.
\newblock {\em Cluster Computing}, 22(3):5905--5917, 2019.

\bibitem{wu2021top}
Youxi Wu, Yuehua Wang, Yan Li, Xingquan Zhu, and Xindong Wu.
\newblock Top-$k$ self-adaptive contrast sequential pattern mining.
\newblock {\em IEEE Transactions on Cybernetics}, 2021.

\bibitem{zheng2016effective}
Zhigang Zheng, Wei Wei, Chunming Liu, Wei Cao, Longbing Cao, and Maninder
  Bhatia.
\newblock An effective contrast sequential pattern mining approach to taxpayer
  behavior analysis.
\newblock {\em World Wide Web}, 19(4):633--651, 2016.

\bibitem{gionis2016bump}
Aristides Gionis, Michael Mathioudakis, and Antti Ukkonen.
\newblock Bump hunting in the dark: Local discrepancy maximization on graphs.
\newblock {\em IEEE Transactions on Knowledge and Data Engineering},
  29(3):529--542, 2016.

\bibitem{jiao2020sub}
Yizhu Jiao, Yun Xiong, Jiawei Zhang, Yao Zhang, Tianqi Zhang, and Yangyong Zhu.
\newblock Sub-graph contrast for scalable self-supervised graph representation
  learning.
\newblock In {\em IEEE International Conference on Data Mining}, pages
  222--231. IEEE, 2020.

\bibitem{ting2006mining}
Roger Ming~Hieng Ting and James Bailey.
\newblock Mining minimal contrast subgraph patterns.
\newblock In {\em {Proceedings of the SIAM International Conference on Data
  Mining}}, pages 639--643. SIAM, 2006.

\bibitem{wang2008spatial}
Bei Wang, Jeff~M Phillips, Robert Schreiber, Dennis Wilkinson, Nina Mishra, and
  Robert Tarjan.
\newblock Spatial scan statistics for graph clustering.
\newblock In {\em Proceedings of the SIAM International Conference on Data
  Mining}, pages 727--738. SIAM, 2008.

\bibitem{yang2018mining}
Yu~Yang, Lingyang Chu, Yanyan Zhang, Zhefeng Wang, Jian Pei, and Enhong Chen.
\newblock Mining density contrast subgraphs.
\newblock In {\em IEEE 34th International Conference on Data Engineering},
  pages 221--232. IEEE, 2018.

\bibitem{li2013efficient}
Jinjiu Li, Can Wang, Wei Wei, Mu~Li, and Chunming Liu.
\newblock Efficient mining of contrast patterns on large scale imbalanced
  real-life data.
\newblock In {\em Pacific-Asia Conference on Knowledge Discovery and Data
  Mining}, pages 62--73. Springer, 2013.

\bibitem{he2019mining}
Zengyou He, Simeng Zhang, Feiyang Gu, and Jun Wu.
\newblock Mining conditional discriminative sequential patterns.
\newblock {\em Information Sciences}, 478:524--539, 2019.

\bibitem{li2017pattern}
Li~Li, Sarah Erfani, and Christopher Leckie.
\newblock A pattern tree based method for mining conditional contrast patterns
  of multi-source data.
\newblock In {\em IEEE International Conference on Data Mining Workshops},
  pages 916--923. IEEE, 2017.

\bibitem{iqbal2017activity}
Mohammad Iqbal and Hsing-Kuo Pao.
\newblock Activity recognition from minimal distinguishing subsequence mining.
\newblock In {\em AIP Conference Proceedings}, volume 1867, page 020046. AIP
  Publishing LLC, 2017.

\bibitem{ji2007mining}
Xiaonan Ji, James Bailey, and Guozhu Dong.
\newblock Mining minimal distinguishing subsequence patterns with gap
  constraints.
\newblock {\em Knowledge and Information Systems}, 11(3):259--286, 2007.

\bibitem{garcia2018overview}
AM~Garc{\'\i}a-Vico, Crist{\'o}bal~J Carmona, Diana Mart{\'\i}n, Milton
  Garc{\'\i}a-Borroto, and Mar{\'\i}a~Jos{\'e} del Jesus.
\newblock An overview of emerging pattern mining in supervised descriptive rule
  discovery: taxonomy, empirical study, trends, and prospects.
\newblock {\em Wiley Interdisciplinary Reviews: Data Mining and Knowledge
  Discovery}, 8(1):e1231, 2018.

\bibitem{loyola2020review}
Octavio Loyola-Gonz{\'a}lez, Miguel~Angel Medina-P{\'e}rez, and
  Kim-Kwang~Raymond Choo.
\newblock A review of supervised classification based on contrast patterns:
  Applications, trends, and challenges.
\newblock {\em Journal of Grid Computing}, pages 1--49, 2020.

\bibitem{dong1999efficient}
Guozhu Dong and Jinyan Li.
\newblock Efficient mining of emerging patterns: Discovering trends and
  differences.
\newblock In {\em The fifth ACM SIGKDD International Conference on Knowledge
  Discovery and Data Mining}, pages 43--52, 1999.

\bibitem{garcia2011fuzzy}
Milton Garc{\'\i}a-Borroto, Jos{\'e}~Fco Mart{\'\i}nez-Trinidad, and
  Jes{\'u}s~Ariel Carrasco-Ochoa.
\newblock Fuzzy emerging patterns for classifying hard domains.
\newblock {\em Knowledge and Information Systems}, 28(2):473--489, 2011.

\bibitem{dong1999discovering}
Guozhu Dong, Jinyan Li, and Xiuzhen Zhang.
\newblock Discovering jumping emerging patterns and experiments on real
  datasets.
\newblock 1999.

\bibitem{wang2004exploiting}
Zhou Wang, Hongjian Fan, and Kotagiri Ramamohanarao.
\newblock Exploiting maximal emerging patterns for classification.
\newblock In {\em Australasian Joint Conference on Artificial Intelligence},
  pages 1062--1068. Springer, 2004.

\bibitem{fan2002efficient}
Hongjian Fan and Ramamohanarao Kotagiri.
\newblock An efficient single-scan algorithm for mining essential jumping
  emerging patterns for classification.
\newblock In {\em Pacific-Asia Conference on Knowledge Discovery and Data
  Mining}, pages 456--462. Springer, 2002.

\bibitem{fan2006fast}
Hongjian Fan and Kotagiri Ramamohanarao.
\newblock Fast discovery and the generalization of strong jumping emerging
  patterns for building compact and accurate classifiers.
\newblock {\em IEEE Transactions on Knowledge and Data Engineering},
  18(6):721--737, 2006.

\bibitem{fan2004noise}
Hongjian Fan and Kotagiri Ramamohanarao.
\newblock Noise tolerant classification by chi emerging patterns.
\newblock In {\em Pacific-Asia Conference on Knowledge Discovery and Data
  Mining}, pages 201--206. Springer, 2004.

\bibitem{chen2016finding}
Xiangtao Chen, Zhouzhou Liu, and SH~Zhu.
\newblock Finding contrast patterns in imbalanced classification based on
  sliding window.
\newblock In {\em The 4th International Conference on Mechanical Materials and
  Manufacturing Engineering, Advances in Engineering Research}, volume~10,
  pages 161--166, 2016.

\bibitem{terlecki2007jumping}
Pawel Terlecki and Krzysztof Walczak.
\newblock Jumping emerging patterns with negation in transaction
  databases--classification and discovery.
\newblock {\em Information Sciences}, 177(24):5675--5690, 2007.

\bibitem{zaiane2007finding}
Osmar~R Za{\i}ane, Kalina Yacef, and Judy Kay.
\newblock Finding top-$n$ emerging sequences to contrast sequence sets.
\newblock Technical report, Department of Computing Science, University of
  Alberta, 2007.

\bibitem{yang2015mining}
Hao Yang, Lei Duan, Bin Hu, Song Deng, W~Wang, and Pan Qin.
\newblock Mining top-$k$ distinguishing sequential patterns with gap
  constraint.
\newblock {\em Journal of Software}, 26(11):2994--3009, 2015.

\bibitem{zhao2015finding}
Yuhai Zhao, Yuan Li, Ying Yin, and Gang Sheng.
\newblock Finding top-covering irreducible contrast sequence rules for disease
  diagnosis.
\newblock {\em Computational and Mathematical Methods in Medicine}, 2015, 2015.

\bibitem{duan2017mining}
Lei Duan, Li~Yan, Guozhu Dong, Jyrki Nummenmaa, and Hao Yang.
\newblock Mining top-$k$ distinguishing temporal sequential patterns from event
  sequences.
\newblock In {\em International Conference on Database Systems for Advanced
  Applications}, pages 235--250. Springer, 2017.

\bibitem{mathonat2021anytime}
Romain Mathonat, Diana Nurbakova, Jean-Fran{\c{c}}ois Boulicaut, and Mehdi
  Kaytoue.
\newblock Anytime mining of sequential discriminative patterns in labeled
  sequences.
\newblock {\em Knowledge and Information Systems}, 63(2):439--476, 2021.

\bibitem{bailey2016statistical}
James Bailey.
\newblock Statistical measures for contrast patterns.
\newblock In {\em {Contrast Data Mining}}, pages 37--44. Chapman and Hall/CRC,
  2016.

\bibitem{liu2015discriminative}
Xiaoqing Liu, Jun Wu, Feiyang Gu, Jie Wang, and Zengyou He.
\newblock Discriminative pattern mining and its applications in bioinformatics.
\newblock {\em Briefings in Bioinformatics}, 16(5):884--900, 2015.

\bibitem{garcia2021cellular}
{\'A}ngel~M Garc{\'\i}a-Vico, Crist{\'o}bal Carmona, Pedro Gonz{\'a}lez, and
  Mar{\'\i}a~J del Jesus.
\newblock A cellular-based evolutionary approach for the extraction of emerging
  patterns in massive data streams.
\newblock {\em Expert Systems with Applications}, 183:115419, 2021.

\bibitem{lucas2017new}
Tarcisio Lucas, T{\'u}lio~CPB Silva, Renato Vimieiro, and Teresa~B Ludermir.
\newblock A new evolutionary algorithm for mining top-$k$ discriminative
  patterns in high dimensional data.
\newblock {\em Applied Soft Computing}, 59:487--499, 2017.

\bibitem{bay1999detecting}
Stephen~D Bay and Michael~J Pazzani.
\newblock Detecting change in categorical data: Mining contrast sets.
\newblock In {\em The 5th ACM SIGKDD International Conference on Knowledge
  Discovery and Data Mining}, pages 302--306, 1999.

\bibitem{gamberger2002expert}
Dragan Gamberger and Nada Lavrac.
\newblock Expert-guided subgroup discovery: Methodology and application.
\newblock {\em Journal of Artificial Intelligence Research}, 17:501--527, 2002.

\bibitem{li2007strong}
Jinyan Li and Qiang Yang.
\newblock Strong compound-risk factors: Efficient discovery through emerging
  patterns and contrast sets.
\newblock {\em IEEE Transactions on Information Technology in Biomedicine},
  11(5):544--552, 2007.

\bibitem{fang2011characterizing}
Gang Fang, Wen Wang, Benjamin Oatley, Brian Van~Ness, Michael Steinbach, and
  Vipin Kumar.
\newblock Characterizing discriminative patterns.
\newblock {\em arXiv preprint, arXiv:1102.4104}, 2011.

\bibitem{garcia2013comparing}
Milton Garc{\'\i}a-Borroto, Octavio Loyola-Gonzalez, Jos{\'e}~Francisco
  Mart{\'\i}nez-Trinidad, and Jes{\'u}s~Ariel Carrasco-Ochoa.
\newblock Comparing quality measures for contrast pattern classifiers.
\newblock In {\em Iberoamerican Congress on Pattern Recognition}, pages
  311--318. Springer, 2013.

\bibitem{bay2001detecting}
Stephen~D Bay and Michael~J Pazzani.
\newblock Detecting group differences: Mining contrast sets.
\newblock {\em Data Mining and Knowledge Discovery}, 5(3):213--246, 2001.

\bibitem{azevedo2010rules}
Paulo~J Azevedo.
\newblock Rules for contrast sets.
\newblock {\em Intelligent Data Analysis}, 14(6):623--640, 2010.

\bibitem{1928Statistical}
Ronald~Aylmer Fisher.
\newblock Statistical methods for research workers.
\newblock In {\em Breakthroughs in statistics}, pages 66--70. Springer, 1992.

\bibitem{None1926The}
Ronald~Aylmer Fisher.
\newblock The goodness of fit of regression formulae, and the distribution of
  regression coefficients.
\newblock {\em Journal of the Royal Statistical Society}, 85(4):597--612, 1922.

\bibitem{garcia2019big}
{\'A}ngel~Miguel Garc{\'\i}a-Vico, Pedro Gonz{\'a}lez, Crist{\'o}bal~Jos{\'e}
  Carmona, and Mar{\'\i}a~Jos{\'e} del Jesus.
\newblock A big data approach for the extraction of fuzzy emerging patterns.
\newblock {\em Cognitive Computation}, 11(3):400--417, 2019.

\bibitem{ramamohanarao2007patterns}
Kotagiri Ramamohanarao and Hongjian Fan.
\newblock Patterns based classifiers.
\newblock {\em World Wide Web}, 10(1):71--83, 2007.

\bibitem{rios2020discovering}
Ingrid~Aylin R{\'\i}os-M{\'e}ndez, Lisbeth Rodr{\'\i}guez-Mazahua, Jos{\'e}
  Antonio~Palet Guzm{\'a}n, Isaac Machorro-Cano, Silvestre~Gustavo
  Pel{\'a}ez-Camarena, Celia Romero-Torres, and Hilarion~Mu{\~n}oz Contreras.
\newblock Discovering emerging patterns from medical opinions about the
  decrease of autopsies performed in a mexican hospital.
\newblock In {\em IEEE 16th International Conference on Automation Science and
  Engineering}, pages 798--803. IEEE, 2020.

\bibitem{piao2016emerging}
Minghao Piao.
\newblock Emerging pattern based prediction of heart diseases and powerline
  safety.
\newblock In {\em Contrast Data Mining}, pages 353--360. Chapman and Hall/CRC,
  2016.

\bibitem{sherhod2013toxicological}
Richard Sherhod, Valerie~J Gillet, Thierry Hanser, Philip~N Judson, and
  Jonathan~D Vessey.
\newblock Toxicological knowledge discovery by mining emerging patterns from
  toxicity data.
\newblock {\em Journal of Cheminformatics}, 5(1):1--1, 2013.

\bibitem{sherhod2012automating}
Richard Sherhod, Valerie~J Gillet, Philip~N Judson, and Jonathan~D Vessey.
\newblock Automating knowledge discovery for toxicity prediction using jumping
  emerging pattern mining.
\newblock {\em Journal of Chemical Information and Modeling},
  52(11):3074--3087, 2012.

\bibitem{sherhod2014emerging}
Richard Sherhod, Philip~N Judson, Thierry Hanser, Jonathan~D Vessey, Samuel~J
  Webb, and Valerie~J Gillet.
\newblock Emerging pattern mining to aid toxicological knowledge discovery.
\newblock {\em Journal of Chemical Information and Modeling}, 54(7):1864--1879,
  2014.

\bibitem{kong2020analysis}
Jie Kong, Jiaxin Han, Junping Ding, Haiyang Xia, and Xin Han.
\newblock Analysis of students’ learning and psychological features by
  contrast frequent patterns mining on academic performance.
\newblock {\em Neural Computing and Applications}, 32(1):205--211, 2020.

\bibitem{thanasuan2017emerging}
Kejkaew Thanasuan, Warasinee Chaisangmongkon, and Chanikarn Wongviriyawong.
\newblock Emerging patterns in student's learning attributes through text
  mining.
\newblock In {\em The 10th International Conference on Educational Data
  Mining}, 2017.

\bibitem{tian2016discovering}
Xianghong Tian, Jie Kong, Tianqing Zhu, and Haiyang Xia.
\newblock Discovering learning patterns of male and female students by contrast
  targeted rule mining.
\newblock In {\em 4th International Conference on Enterprise Systems}, pages
  196--202. IEEE, 2016.

\bibitem{dong2018oclep+}
Guozhu Dong and Sai~Kiran Pentukar.
\newblock {OCLEP+}: One-class anomaly and intrusion detection using minimal
  length of emerging patterns.
\newblock {\em arXiv preprint, arXiv:1811.09842}, 2018.

\bibitem{hellal2016minimal}
Aya Hellal and Lotfi~Ben Romdhane.
\newblock Minimal contrast frequent pattern mining for malware detection.
\newblock {\em Computers \& Security}, 62:19--32, 2016.

\bibitem{dong2009applications}
Guozhu Dong and Jinyan Li.
\newblock Applications of emerging patterns for microarray gene expression data
  analysis.
\newblock {\em Encyclopedia of Database Systems, Second Edition}, 2009.

\bibitem{namasivayam2013prediction}
Vigneshwaran Namasivayam, Preeti Iyer, and Jurgen Bajorath.
\newblock Prediction of individual compounds forming activity cliffs using
  emerging chemical patterns.
\newblock {\em Journal of Chemical Information and Modeling},
  53(12):3131--3139, 2013.

\bibitem{garcia2016analysing}
AM~Garc{\'\i}a-Vico, J~Montes, Jorge Aguilera, Crist{\'o}bal~J Carmona, and
  Mar{\'\i}a~Jos{\'e} del Jesus.
\newblock Analysing concentrating photovoltaics technology through the use of
  emerging pattern mining.
\newblock In {\em International Joint Conference SOCO-CISIS-ICEUTE}, pages
  334--344. Springer, 2016.

\bibitem{weng2018observation}
Cheng-Hsiung Weng and Cheng-Kui~Huang Tony.
\newblock Observation of sales trends by mining emerging patterns in dynamic
  markets.
\newblock {\em Applied Intelligence}, 48(11):4515--4529, 2018.

\bibitem{li2004deeps}
Jinyan Li, Guozhu Dong, Kotagiri Ramamohanarao, and Limsoon Wong.
\newblock {DeEPs}: A new instance-based lazy discovery and classification
  system.
\newblock {\em Machine Learning}, 54(2):99--124, 2004.

\bibitem{li2000space}
Jinyan Li, Kotagiri Ramamohanarao, and Guozhu Dong.
\newblock The space of jumping emerging patterns and its incremental
  maintenance algorithms.
\newblock In {\em The Seventeenth International Conference on Machine
  Learning}, pages 551--558, 2000.

\bibitem{terlecki2008efficient}
Pawel Terlecki and Krzysztof Walczak.
\newblock Efficient discovery of top-$k$ minimal jumping emerging patterns.
\newblock In {\em International Conference on Rough Sets and Current Trends in
  Computing}, pages 438--447. Springer, 2008.

\bibitem{fan2003efficiently}
Hongjian Fan and Kotagiri Ramamohanarao.
\newblock Efficiently mining interesting emerging patterns.
\newblock In {\em International Conference on Web-Age Information Management},
  pages 189--201. Springer, 2003.

\bibitem{liu2014novel}
Quanzhong Liu, Peng Shi, Zhengguo Hu, and Yang Zhang.
\newblock {A novel approach of mining strong jumping emerging patterns based on
  BSC-tree}.
\newblock {\em International Journal of Systems Science}, 45(3):598--615, 2014.

\bibitem{fan2003bayesian}
Hongjian Fan and Kotagiri Ramamohanarao.
\newblock A bayesian approach to use emerging patterns for classification.
\newblock In {\em The 14th Australasian Database Conference}, pages 39--48,
  2003.

\bibitem{li2019mining}
Qing Li, Xiangtao Chen, and Ronghui Wu.
\newblock Mining contrast sequential patterns based on subsequence location
  distribution from biological sequences.
\newblock In {\em The 2nd International Conference on Data Science and
  Information Technology}, pages 204--209, 2019.

\bibitem{wang2014efficient}
Xianming Wang, Lei Duan, Guozhu Dong, Zhonghua Yu, and Changjie Tang.
\newblock Efficient mining of density-aware distinguishing sequential patterns
  with gap constraints.
\newblock In {\em International Conference on Database Systems for Advanced
  Applications}, pages 372--387. Springer, 2014.

\bibitem{bailey2002fast}
James Bailey, Thomas Manoukian, and Kotagiri Ramamohanarao.
\newblock Fast algorithms for mining emerging patterns.
\newblock In {\em European Conference on Principles of Data Mining and
  Knowledge Discovery}, pages 39--50. Springer, 2002.

\bibitem{li2007mining}
Jinyan Li, Guimei Liu, and Limsoon Wong.
\newblock Mining statistically important equivalence classes and
  delta-discriminative emerging patterns.
\newblock In {\em The 13th ACM SIGKDD International Conference on Knowledge
  Discovery and Data Mining}, pages 430--439, 2007.

\bibitem{piao2011enumeration}
Minghao Piao, Jong~Bum Lee, Ho~Sun Shon, Unil Yun, and Keun~Ho Ryu.
\newblock Enumeration tree based emerging patterns mining by using two
  different supports.
\newblock In {\em International Conference on Hybrid Information Technology},
  pages 708--715. Springer, 2011.

\bibitem{garcia2010lcmine}
Milton Garc{\'\i}a-Borroto, Jos{\'e}~Fco Mart{\'\i}nez-Trinidad,
  Jes{\'u}s~Ariel Carrasco-Ochoa, Miguel~Angel Medina-P{\'e}rez, and Jos{\'e}
  Ruiz-Shulcloper.
\newblock {LCMine}: An efficient algorithm for mining discriminative
  regularities and its application in supervised classification.
\newblock {\em Pattern Recognition}, 43(9):3025--3034, 2010.

\bibitem{garcia2010new}
Milton Garc{\'\i}a-Borroto, Jos{\'e}~Francisco Mart{\'\i}nez-Trinidad, and
  Jes{\'u}s~Ariel Carrasco-Ochoa.
\newblock A new emerging pattern mining algorithm and its application in
  supervised classification.
\newblock In {\em Pacific-Asia Conference on Knowledge Discovery and Data
  Mining}, pages 150--157. Springer, 2010.

\bibitem{wang2010building}
Liang Wang, Yizhou Wang, and Debin Zhao.
\newblock {Building emerging pattern (EP) random forest for recognition}.
\newblock In {\em IEEE International Conference on Image Processing}, pages
  1457--1460. IEEE, 2010.

\bibitem{M2015Finding}
M~García-Borroto, JF~Martínez-Trinidad, and J.~A. Carrasco-Ochoa.
\newblock Finding the best diversity generation procedures for mining contrast
  patterns.
\newblock {\em Expert Systems with Applications}, 42(11), 2015.

\bibitem{2017PBC4cip}
Octavio Loyola-Gonzalez, Miguel Angel Medina-Perez, Jose~Fco.
  Martinez-Trinidad, Jesus Ariel Carrasco-Ochoa, Raul Monroy, and Milton
  Garcia-Borroto.
\newblock {PBC4Cip}: A new contrast pattern-based classifier for class
  imbalance problems.
\newblock {\em Knowledge-Based Systems}, 115(JAN.1):100--109, 2017.

\bibitem{canete2019classification}
Leonardo Ca{\~n}ete-Sifuentes, Ra{\'u}l Monroy, Miguel~Angel Medina-P{\'e}rez,
  Octavio Loyola-Gonz{\'a}lez, and Francisco~Vera Voronisky.
\newblock Classification based on multivariate contrast patterns.
\newblock {\em IEEE Access}, 7:55744--55762, 2019.

\bibitem{garcia2018moea}
{\'A}ngel~Miguel Garc{\'\i}a-Vico, Crist{\'o}bal~Jos{\'e} Carmona, Pedro
  Gonz{\'a}lez, and Mar{\'\i}a~Jos{\'e} del Jesus.
\newblock {MOEA-EFEP}: Multi-objective evolutionary algorithm for extracting
  fuzzy emerging patterns.
\newblock {\em IEEE Transactions on Fuzzy Systems}, 26(5):2861--2872, 2018.

\bibitem{garcia2017first}
AM~Garc{\'\i}a-Vico, Pedro Gonz{\'a}lez, Mar{\'\i}a~Jos{\'e} del Jesus, and
  Crist{\'o}bal~J Carmona.
\newblock A first approach to handle fuzzy emerging patterns mining on big data
  problems: The {EvAEFP-spark} algorithm.
\newblock In {\em IEEE International Conference on Fuzzy Systems}, pages 1--6.
  IEEE, 2017.

\bibitem{carmona2018unifying}
Crist{\'o}bal~J Carmona, Mar{\'\i}a~Jos{\'e} del Jesus, and Francisco Herrera.
\newblock A unifying analysis for the supervised descriptive rule discovery via
  the weighted relative accuracy.
\newblock {\em Knowledge-Based Systems}, 139:89--100, 2018.

\bibitem{garcia2020e2pamea}
Angel~Miguel Garcia-Vico, Francisco Charte, Pedro Gonzalez, David Elizondo, and
  Cristobal~Jose Carmona.
\newblock {E2PAMEA: A fast evolutionary algorithm for extracting fuzzy emerging
  patterns in big data environments}.
\newblock {\em Neurocomputing}, 415:60--73, 2020.

\bibitem{garcia2014survey}
Milton Garc{\'\i}a-Borroto, Jos{\'e}~Fco Mart{\'\i}nez-Trinidad, and
  Jes{\'u}s~Ariel Carrasco-Ochoa.
\newblock A survey of emerging patterns for supervised classification.
\newblock {\em Artificial Intelligence Review}, 42(4):705--721, 2014.

\bibitem{loekito2006fast}
Elsa LoekitoA Review~of Supervised and James Bailey.
\newblock Fast mining of high dimensional expressive contrast patterns using
  zero-suppressed binary decision diagrams.
\newblock In {\em The 12th ACM SIGKDD International Conference on Knowledge
  Discovery and Data Mining}, pages 307--316, 2006.

\bibitem{gao2016mining}
Chao Gao, Lei Duan, Guozhu Dong, Haiqing Zhang, Hao Yang, and Changjie Tang.
\newblock Mining top-$k$ distinguishing sequential patterns with flexible gap
  constraints.
\newblock In {\em International Conference on Web-Age Information Management},
  pages 82--94. Springer, 2016.

\bibitem{ferreira2001gene}
Candida Ferreira.
\newblock Gene expression programming: a new adaptive algorithm for solving
  problems.
\newblock {\em arXiv preprint, cs/0102027}, 2001.

\bibitem{loyola2019cost}
Octavio Loyola-Gonz{\'a}lez, Jos{\'e}~FCO Mart{\'\i}nez-Trinidad,
  Jes{\'u}s~Ariel Carrasco-Ochoa, and Milton Garcia-Borroto.
\newblock Cost-sensitive pattern-based classification for class imbalance
  problems.
\newblock {\em IEEE Access}, 7:60411--60427, 2019.

\bibitem{krawczyk2016learning}
Bartosz Krawczyk.
\newblock Learning from imbalanced data: open challenges and future directions.
\newblock {\em Progress in Artificial Intelligence}, 5(4):221--232, 2016.

\bibitem{ng2021maximizing}
Wing~WY Ng, Zhengxi Liu, Jianjun Zhang, and Witold Pedrycz.
\newblock Maximizing minority accuracy for imbalanced pattern classification
  problems using cost-sensitive localized generalization error model.
\newblock {\em Applied Soft Computing}, 104:107178, 2021.

\bibitem{garcia2020fepds}
{\'A}ngel~Miguel Garc{\'\i}a-Vico, Crist{\'o}bal~J Carmona, Pedro Gonz{\'a}lez,
  Huseyin Seker, and Mar{\'\i}a~J del Jesus.
\newblock {FEPDS}: A proposal for the extraction of fuzzy emerging patterns in
  data streams.
\newblock {\em IEEE Transactions on Fuzzy Systems}, 28(12):3193--3203, 2020.

\bibitem{savage2016distributed}
David Savage, Xiuzhen Zhang, Pauline Chou, Xinghou Yu, and Qingmai Wang.
\newblock Distributed mining of contrast patterns.
\newblock {\em IEEE Transactions on Parallel and Distributed Systems},
  28(7):1881--1890, 2016.

\bibitem{2018MapReduce}
Jeffrey Dean and Sanjay Ghemawat.
\newblock {MapReduce}: Simplified data processing on large clusters.
\newblock {\em Communications of the ACM}, 51(1):107--113, 2008.

\bibitem{zaharia2010spark}
Matei Zaharia, Mosharaf Chowdhury, Michael~J Franklin, Scott Shenker, and Ion
  Stoica.
\newblock {Spark}: Cluster computing with working sets.
\newblock In {\em The 2nd USENIX Workshop on Hot Topics in Cloud Computing},
  2010.

\bibitem{wang2015contrast}
Dawei Wang.
\newblock Contrast pattern based methods for visualizing and predicting
  spatiotemporal events.
\newblock In {\em IEEE International Conference on Data Mining Workshop}, pages
  1560--1567. IEEE, 2015.

\bibitem{nishiguchi2016caecp}
Mao Nishiguchi and Hiroyuki Morita.
\newblock {CAECP and CRPD: Classification by aggregating essential contrast
  patterns, and contrast ranked path diagrams}.
\newblock {\em {Journal of Information \& Knowledge Management}},
  15(04):1650045, 2016.

\bibitem{loyola2020contrast}
Octavio Loyola-Gonz{\'a}lez, Miguel~Angel Medina-P{\'e}rez, Raymundo
  Adri{\'a}n~Coronilla Valdez, and Kim-Kwang~Raymond Choo.
\newblock A contrast pattern-based scientometric study of the {QS} world
  university ranking.
\newblock {\em IEEE Access}, 8:206088--206104, 2020.

\bibitem{neto2021multivariate}
M{\'a}rio~Popolin Neto and Fernando~V Paulovich.
\newblock {Multivariate data explanation by jumping emerging patterns
  visualization}.
\newblock {\em arXiv preprint, arXiv:2106.11112}, 2021.

\bibitem{gubbi2013internet}
Jayavardhana Gubbi, Rajkumar Buyya, Slaven Marusic, and Marimuthu Palaniswami.
\newblock Internet of things {(IoT)}: A vision, architectural elements, and
  future directions.
\newblock {\em Future Generation Computer Systems}, 29(7):1645--1660, 2013.

\bibitem{rappaport2013millimeter}
Theodore~S Rappaport, Shu Sun, Rimma Mayzus, Hang Zhao, Yaniv Azar, Kevin Wang,
  George~N Wong, Jocelyn~K Schulz, Mathew Samimi, and Felix Gutierrez.
\newblock Millimeter wave mobile communications for {5G} cellular: It will
  work!
\newblock {\em IEEE Access}, 1:335--349, 2013.

\bibitem{wang2010cloud}
Lizhe Wang, Gregor Von~Laszewski, Andrew Younge, Xi~He, Marcel Kunze, Jie Tao,
  and Cheng Fu.
\newblock Cloud computing: a perspective study.
\newblock {\em New Generation Computing}, 28(2):137--146, 2010.

\bibitem{swan2015blockchain}
Melanie Swan.
\newblock {\em Blockchain: Blueprint for a new economy}.
\newblock " O'Reilly Media, Inc.", 2015.

\bibitem{2021Artificial}
M.~M. Mijwil and R.~A. Abttan.
\newblock Artificial intelligence: A survey on evolution and future trends.
\newblock {\em Asian Journal of Applied Sciences}, 9(2):87--93, 2021.

\bibitem{ahmad2019security}
Ijaz Ahmad, Shahriar Shahabuddin, Tanesh Kumar, Jude Okwuibe, Andrei Gurtov,
  and Mika Ylianttila.
\newblock Security for 5g and beyond.
\newblock {\em IEEE Communications Surveys \& Tutorials}, 21(4):3682--3722,
  2019.

\bibitem{sunhare2020internet}
Priyank Sunhare, Rameez~R Chowdhary, and Manju~K Chattopadhyay.
\newblock Internet of things and data mining: An application oriented survey.
\newblock {\em Journal of King Saud University-Computer and Information
  Sciences}, 2020.

\bibitem{shadroo2018systematic}
Shabnam Shadroo and Amir~Masoud Rahmani.
\newblock Systematic survey of big data and data mining in internet of things.
\newblock {\em Computer Networks}, 139:19--47, 2018.

\bibitem{gaber2019internet}
Mohamed~Medhat Gaber, Adel Aneiba, Shadi Basurra, Oliver Batty, Ahmed~M
  Elmisery, Yevgeniya Kovalchuk, and Muhammad Habib~Ur Rehman.
\newblock Internet of things and data mining: From applications to techniques
  and systems.
\newblock {\em Wiley Interdisciplinary Reviews: Data Mining and Knowledge
  Discovery}, 9(3):e1292, 2019.

\bibitem{idhammad2018distributed}
Mohamed Idhammad, Karim Afdel, and Mustapha Belouch.
\newblock Distributed intrusion detection system for cloud environments based
  on data mining techniques.
\newblock {\em Procedia Computer Science}, 127:35--41, 2018.

\bibitem{malazi2018combining}
Hadi~Tabatabaee Malazi and Mohammad Davari.
\newblock Combining emerging patterns with random forest for complex activity
  recognition in smart homes.
\newblock {\em Applied Intelligence}, 48(2):315--330, 2018.

\bibitem{akhriza2015novel}
Tubagus~M Akhriza, Yinghua Ma, and Jianhua Li.
\newblock {A novel Fibonacci windows model for finding emerging patterns over
  online data stream}.
\newblock In {\em International Conference on Cyber Security of Smart Cities,
  Industrial Control System and Communications}, pages 1--8. IEEE, 2015.

\bibitem{bailey2010efficient}
James Bailey and Elsa Loekito.
\newblock Efficient incremental mining of contrast patterns in changing data.
\newblock {\em Information Processing Letters}, 110(3):88--92, 2010.

\bibitem{van2015dynamically}
Jan~Sipke Van Der~Veen, Bram Van Der~Waaij, Elena Lazovik, Wilco Wijbrandi, and
  Robert~J Meijer.
\newblock Dynamically scaling apache storm for the analysis of streaming data.
\newblock In {\em IEEE First International Conference on Big Data Computing
  Service and Applications}, pages 154--161. IEEE, 2015.

\bibitem{2016DPcode}
Zhan Qin, Kui Ren, Ting Yu, and Jian Weng.
\newblock {DPcode}: Privacy-preserving frequent visual patterns publication on
  cloud.
\newblock {\em IEEE Transactions on Multimedia}, 18(5):929--939, 2016.

\bibitem{2018Privacy}
Carson~K Leung, Calvin~SH Hoi, Adam~GM Pazdor, Bryan~H Wodi, and Alfredo
  Cuzzocrea.
\newblock Privacy-preserving frequent pattern mining from big uncertain data.
\newblock In {\em IEEE International Conference on Big Data}, pages 5101--5110.
  IEEE, 2018.

\bibitem{zhang2017efficient}
Yaling Zhang, Ting Wang, and Shangping Wang.
\newblock An efficient algorithm for frequent pattern mining based on
  privacy-preserving.
\newblock In {\em 13th International Conference on Natural Computation, Fuzzy
  Systems and Knowledge Discovery}, pages 1694--1699. IEEE, 2017.

\bibitem{chuang2020debiased}
Ching-Yao Chuang, Joshua Robinson, Yen-Chen Lin, Antonio Torralba, and Stefanie
  Jegelka.
\newblock Debiased contrastive learning.
\newblock {\em Advances in neural information processing systems},
  33:8765--8775, 2020.

\bibitem{khosla2020supervised}
Prannay Khosla, Piotr Teterwak, Chen Wang, Aaron Sarna, Yonglong Tian, Phillip
  Isola, Aaron Maschinot, Ce~Liu, and Dilip Krishnan.
\newblock Supervised contrastive learning.
\newblock {\em Advances in Neural Information Processing Systems},
  33:18661--18673, 2020.

\end{thebibliography}

\end{document}